\documentclass[aps,pra,showpacs,twocolumn,superscriptaddress]{revtex4-1}
\usepackage{amssymb}
\usepackage{amsmath}
\usepackage{graphicx}
\usepackage{epsfig}
\usepackage{subfigure}

\begin{document}

\title{Simple harmonic oscillation in non-Hermitian SSH chain at exceptional
point}
\author{K. L. Zhang}
\affiliation{School of Physics, Nankai University, Tianjin 300071,
	China}
\author{P. Wang}
\affiliation{School of Physics, Nankai University, Tianjin 300071,
	China}
\author{G. Zhang}
\affiliation{College of Physics and Materials Science, Tianjin Normal University, Tianjin
	300387, China}
\author{Z. Song}
\email{songtc@nankai.edu.cn}
\affiliation{School of Physics, Nankai University, Tianjin 300071,
	China}

\begin{abstract}
The balance of gain and loss in an open system may maintain certain
Hermitian dynamical behaviors, which can be hardly observed in a popular Hermitian system. In this paper,
we systematically study a 1D $\mathcal{PT}$-symmetry non-Hermitian SSH model
with open boundary condition based on exact approximate solution. We show
that the long-wave length standing-wave modes can be achieved within the
linear dispersion region when the system is tuned at the exceptional point
(EP). The whole Hilbert space can be decomposed into two quasi-Hermitian
subspaces, which are consisted of positive and negative energy levels,
respectively. Within each subspace, the system maintains all the features of
a Hermitian one. We construct a coherent-like state in a subspace and find
that it exhibits perfect simple harmonic motion (SHM). In contrast to a
canonical coherent state, the shape of the wavepacket deforms periodically
rather than entirely translation. And the amplitude of the SHM is not
determined by the initial condition but the shape of the wavepacket. Our
result indicates that novel Hermitian dynamics can be realized by a
non-Hermitian system.
\end{abstract}

\maketitle

%\pacs{11.30.Er, 03.65.-w, 03.75.-b}

% Force line breaks with \\

\section{Introduction}

\label{sec.intro}

The quantum harmonic oscillator as the quantum-mechanical analog of the
classical harmonic oscillator is one of the most important model systems in
quantum mechanics, not only due to its broad application, but also one of
the few exactly solvable quantum system \cite{Griffiths,Liboff,Zeng}.
Coherent states (also called Glauber states) take an important role not
only to the connection between quantum mechanics and classical mechanics
\cite{Glauber}, but also to the applications in many aspects.

Recently, non-Hermitian quantum mechanics \cite{Bender 98,Bender 99,Dorey
01,Dorey 02,Bender 02,A.M43,A.M36,Jones,M.Z40,M.Z41,M.Z82} attracted
increasing interest in several branches of physics \cite%
{AGuo,CERuter,Wan,Sun,LFeng,BPeng,LChang,LFengScience,HodaeiScience,NC2015}.
Due to the reality of the spectrum in a non-Hermitian system, the connection
to a Hermitian system is an interesting topic. A metric-operator method has
been proposed to compose a Hermitian Hamiltonian, which has exactly the same
real spectrum with the pseudo-Hermitian Hamiltonian \cite{A.M}. From the
Hermitian counterpart, one can extract the physical meaning of a
pseudo-Hermitian Hamiltonian in\ the viewpoint of spectrum \cite%
{A.M38,A.M391,A.M392,JLPT}. Alternatively, a connection between a
non-Hermitian Hamiltonian and an infinite Hermitian system can be
established in the viewpoint of eigen state \cite{L.J81,L.J83,L.J44,LJin84}.
It is found that a non-Hermitian system can exhibit conditional Hermitian
dynamical behaviors, which preserves the Dirac probability due to the
balance of gain and loss. Such dynamics is hard to realized in a non-trivial
Hermitian system. Therefore, it provides a new way to prepare and control
the quantum states in the field of quantum information and technology.

In this paper we present a pseudo-Hermitian system to demonstrate that a
coherent state can be formed in a discrete system by introducing $\mathcal{PT%
}$-symmetry staggered imaginary potentials rather than a real parabolic
potential as studied in Refs.\cite{ST}. The system is described by a 1D $%
\mathcal{PT}$-symmetry non-Hermitian SSH model with open boundary condition.
The exact solution is obtained in the strong dimerization limit, which shows
that the equal-level-spacing high-frequency standing-wave modes (EHSM) can
be achieved when the system is tuned at the exceptional point (EP). The
gapless system supports two quasi-Hermitian subspaces, which are consisted
of positive and negative energy levels, respectively. A coherent state
constructed in a subspace exhibits perfect simple harmonic motion (SHM). In
contrast to a canonical coherent state, the shape of the wavepacket deforms
periodically rather than entirely translation. And the amplitude of the SHM
is not determined by the initial condition but the shape of the wavepacket.
Our result indicates that novel Hermitian dynamics can be realized by a
non-Hermitian system.

The remainder of this paper is organized as follows. In Sec. \ref{Model and
solution}, we present a non-Hermitian\ SSH chain model and the formulation
of approximate diagonalization. Sec. \ref{Hermitian dynamics in sub-space}
reveals the Hermitian dynamics in sub-spaces with positive and negative
eigen energies. Sec. \ref{Simple harmonic oscillation} demonstrates a
peculiar quantum state living in such a non-Hermitian gapless system.
Finally, we present a summary and discussion in Sec. \ref{Summary}.

\section{Model and solution}

\label{Model and solution}

\begin{figure}[tbp]
\centering
\includegraphics[ bb=10 260 650 760, width=0.5\textwidth, clip]{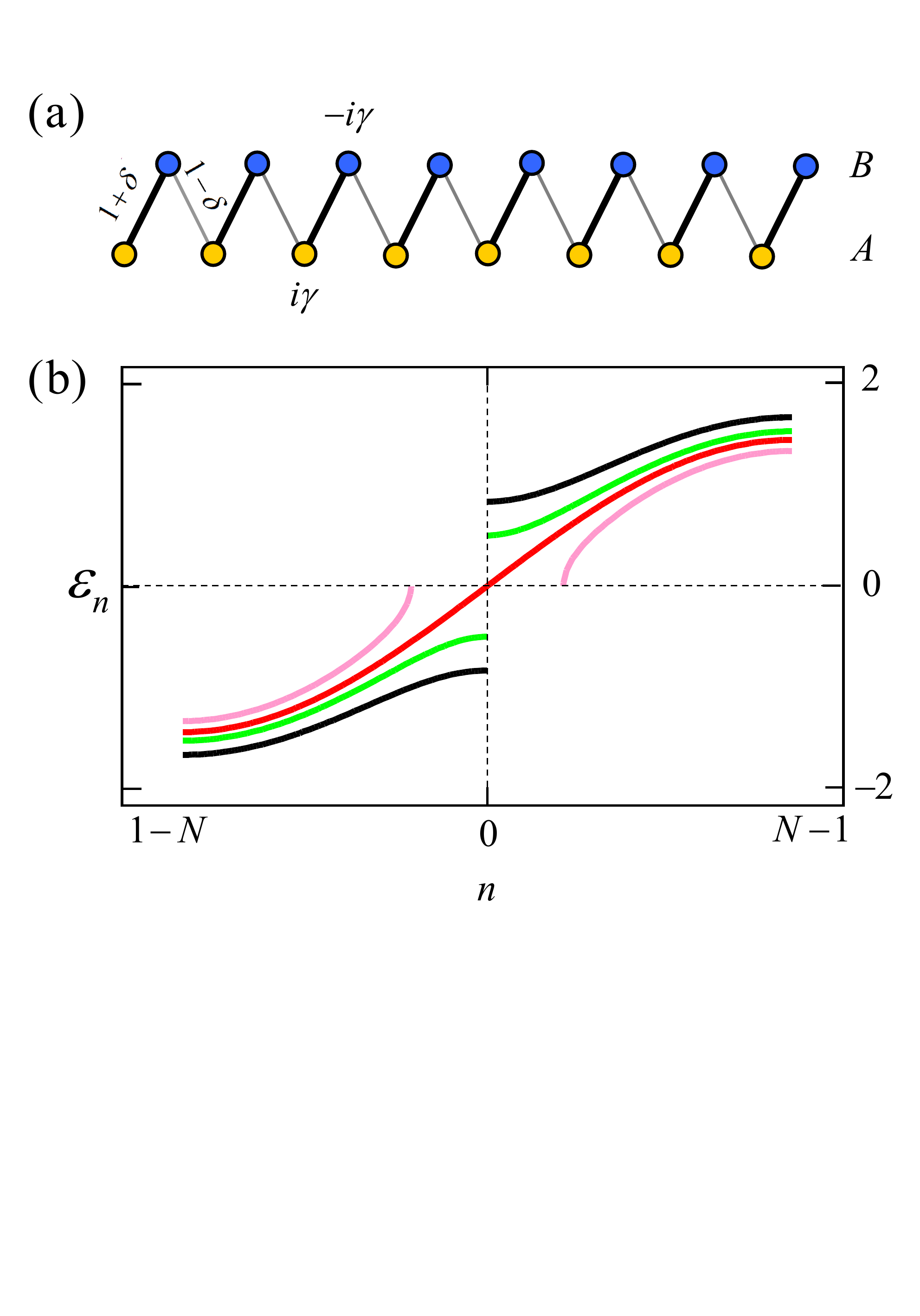}
\caption{(Color online) (a) Schematic for the non-Hermitian SSH chain with $%
A$ (golden) and $B$ (blue) sub-lattices. Black thick and gray thin lines
indicate the hoppping between two nearest neighbor sites with amplitudes $%
\left( 1+\protect\delta \right) $\ and $\left( 1-\protect\delta \right) $,
respectively. Two sub-lattices have opposite site imaginary potentials $\pm i%
\protect\gamma $, representing the physical gain and loss. It has $\mathcal{%
PT}$-symmetry and can have full real spectrum in the case of $1>\protect%
\delta >0$\ (see text). (b) The spectra of the SSH chains with typical
values of $\protect\gamma =0$ (black), $0.8$ (green), $1.0$ (red), and $1.2$
(pink), obtained by exact diagonalization for system with large $N$. We take
$\protect\delta =0.5$, which rules out the mid-gap levels in the case of $%
\protect\gamma =0.0$. There are $2N$ levels for gapped spectra. The gap
closes for $\protect\gamma =1.0$, at which two levels with $n=0$ coalescence
at zero energy, supporting gapless real spectrum with $2N-1$ levels. It
indicates that the energy level spacing is identical for small $n$. For $%
\protect\gamma =1.2$, the imaginary energy levels appear in association with
the $\mathcal{PT}$-symmetry breaking.}
\label{Fig1}
\end{figure}

\begin{figure*}[!t]
\centering
\subfigure[$\ \protect\alpha =\left| \alpha \right|$]
{\includegraphics[height=2in,width=2.2in]{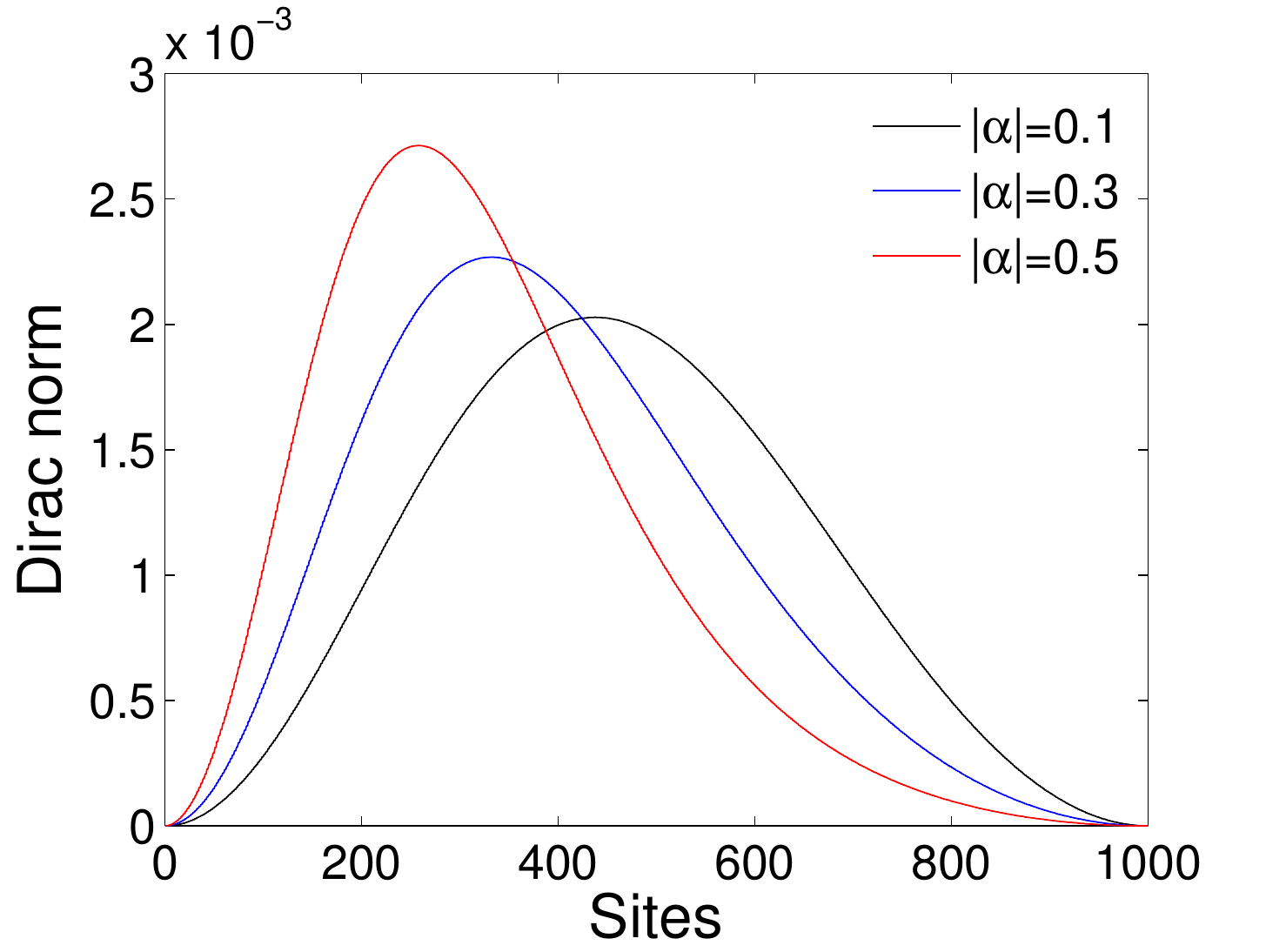}}
\subfigure[$\
	\protect\alpha = \left| \alpha \right|\exp \left(\frac{i \pi }{2}\right)$] {\includegraphics[height=2in,width=2.2in]{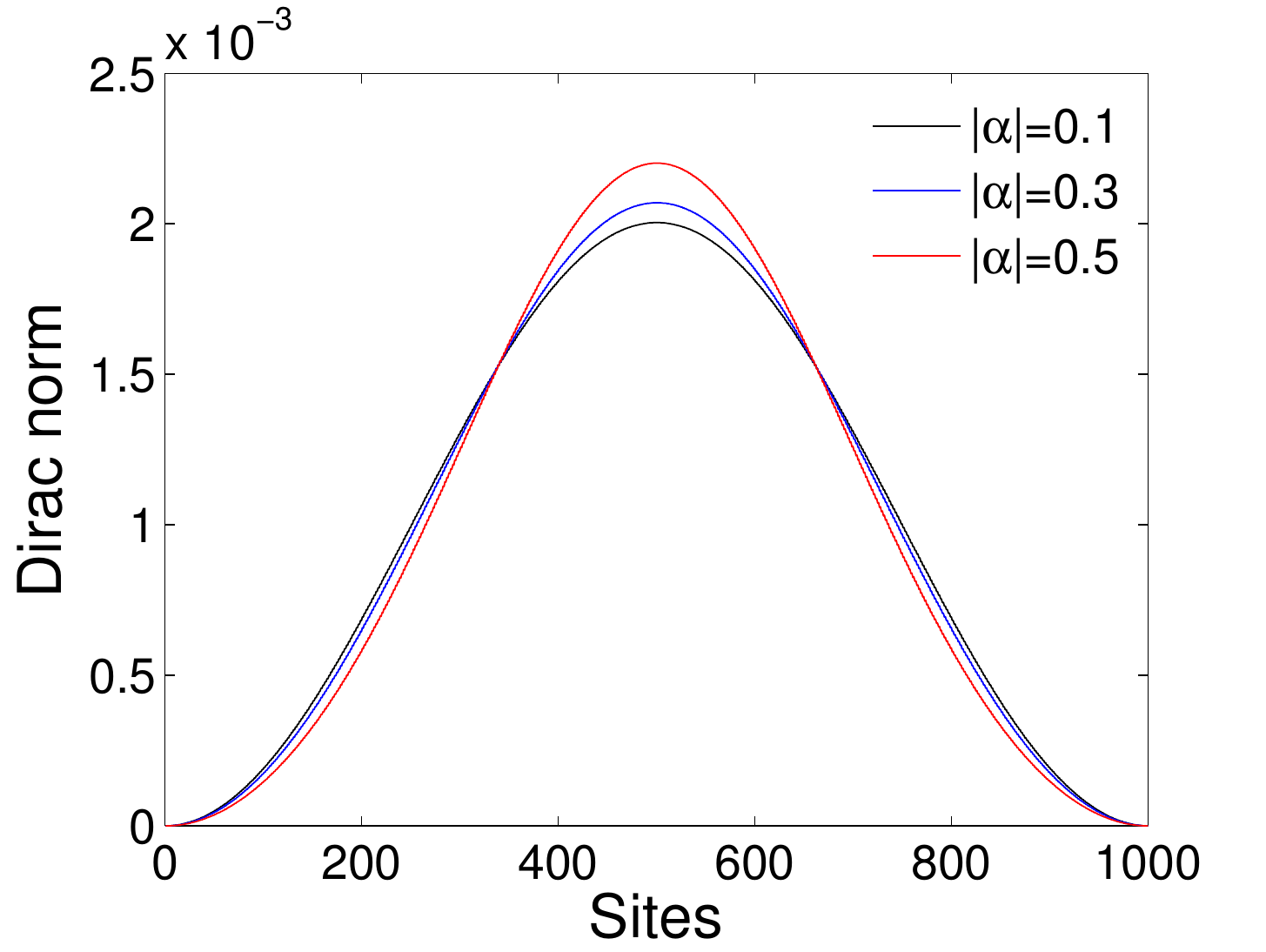}}
\subfigure[$\
	\protect\alpha = \left| \alpha \right|\exp (i \pi )$] {\includegraphics[height=2in,width=2.2in]{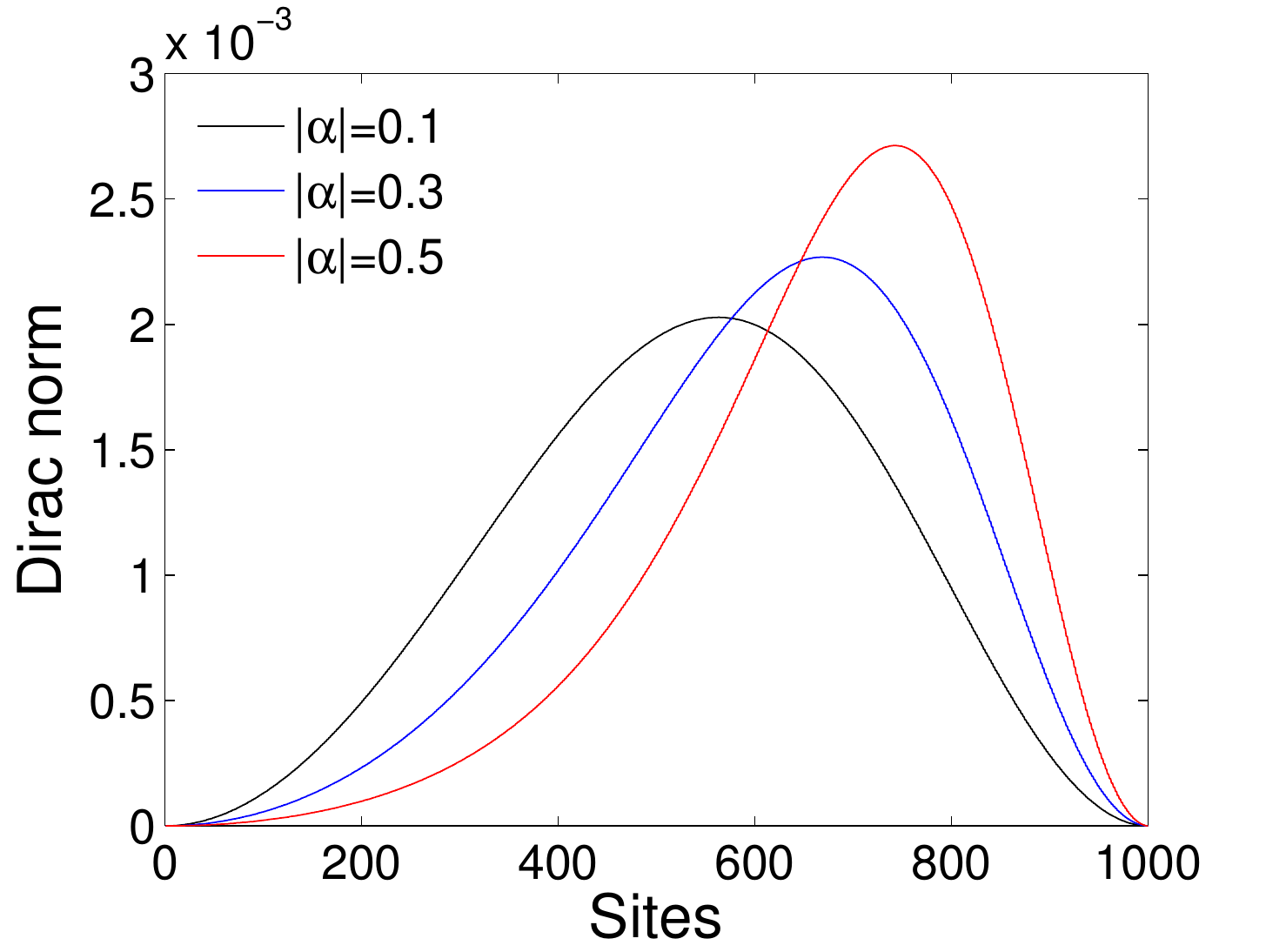}}
\subfigure[$\ \protect\alpha =\left| \alpha \right|$]
{\includegraphics[height=2in,width=2.2in]{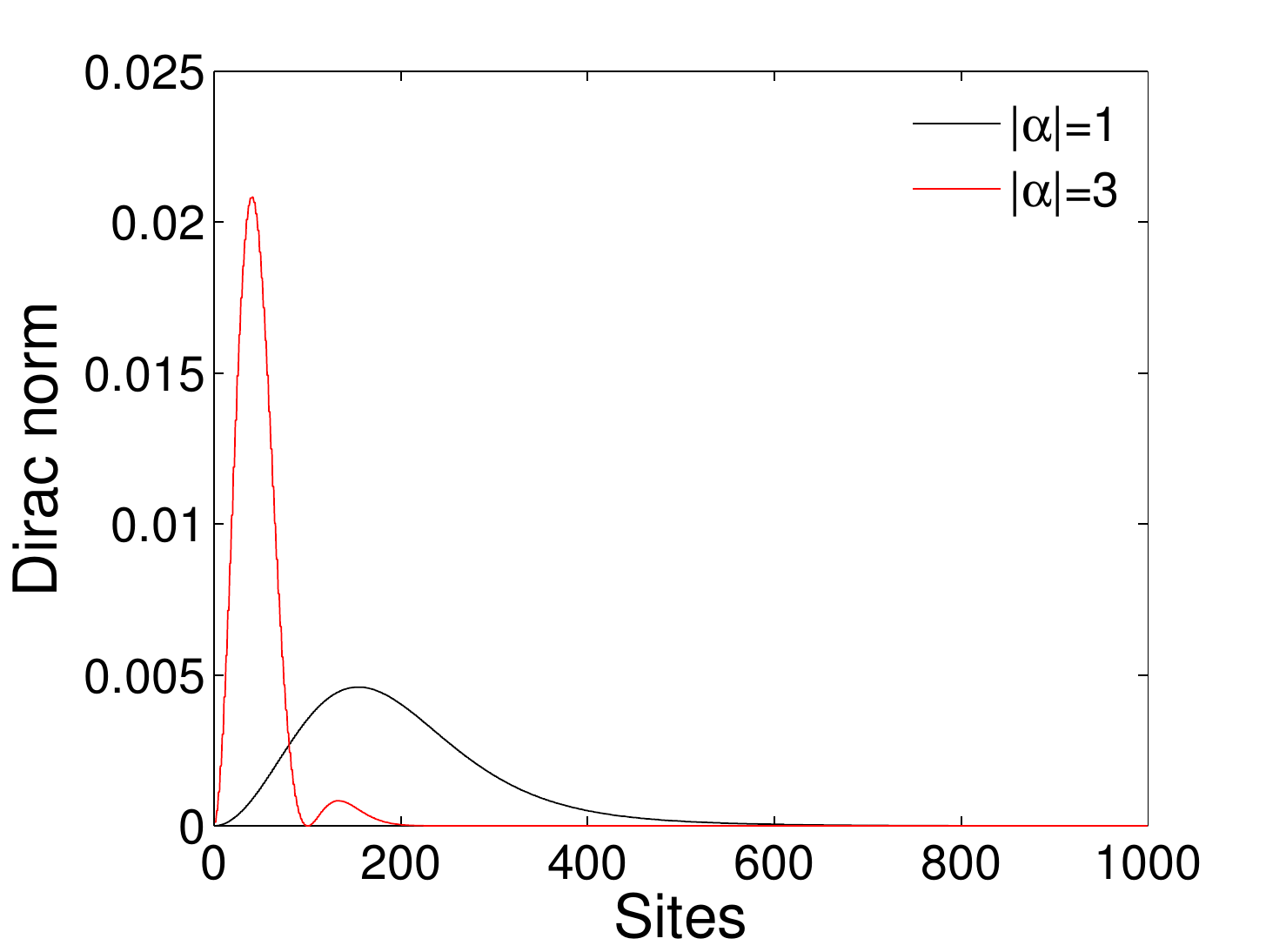}}
\subfigure[$\
\protect\alpha = \left| \alpha \right|\exp \left(\frac{i \pi }{2}\right)$] {\includegraphics[height=2in,width=2.2in]{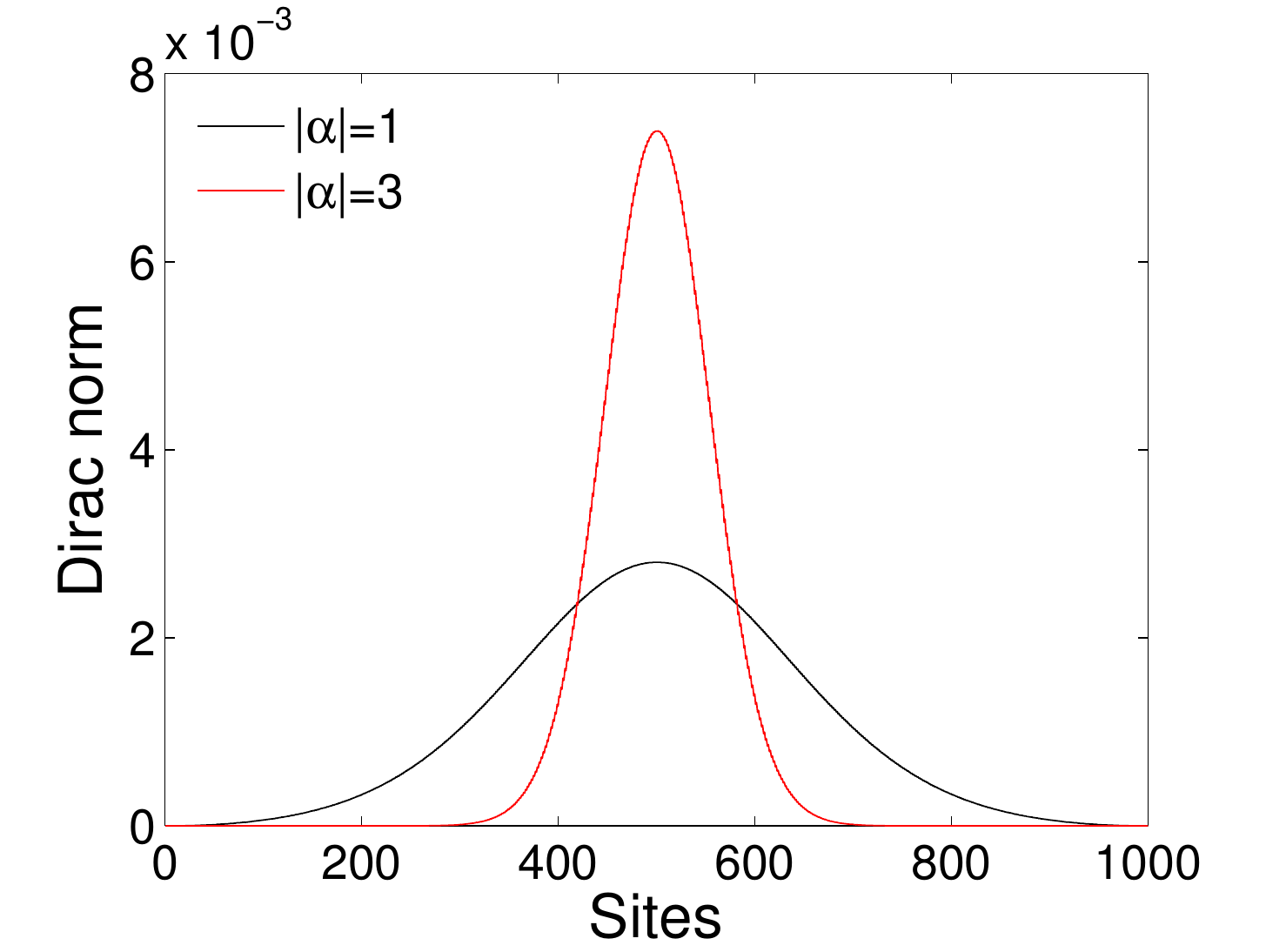}}
\subfigure[$\
\protect\alpha = \left| \alpha \right|\exp (i \pi )$] {\includegraphics[height=2in,width=2.2in]{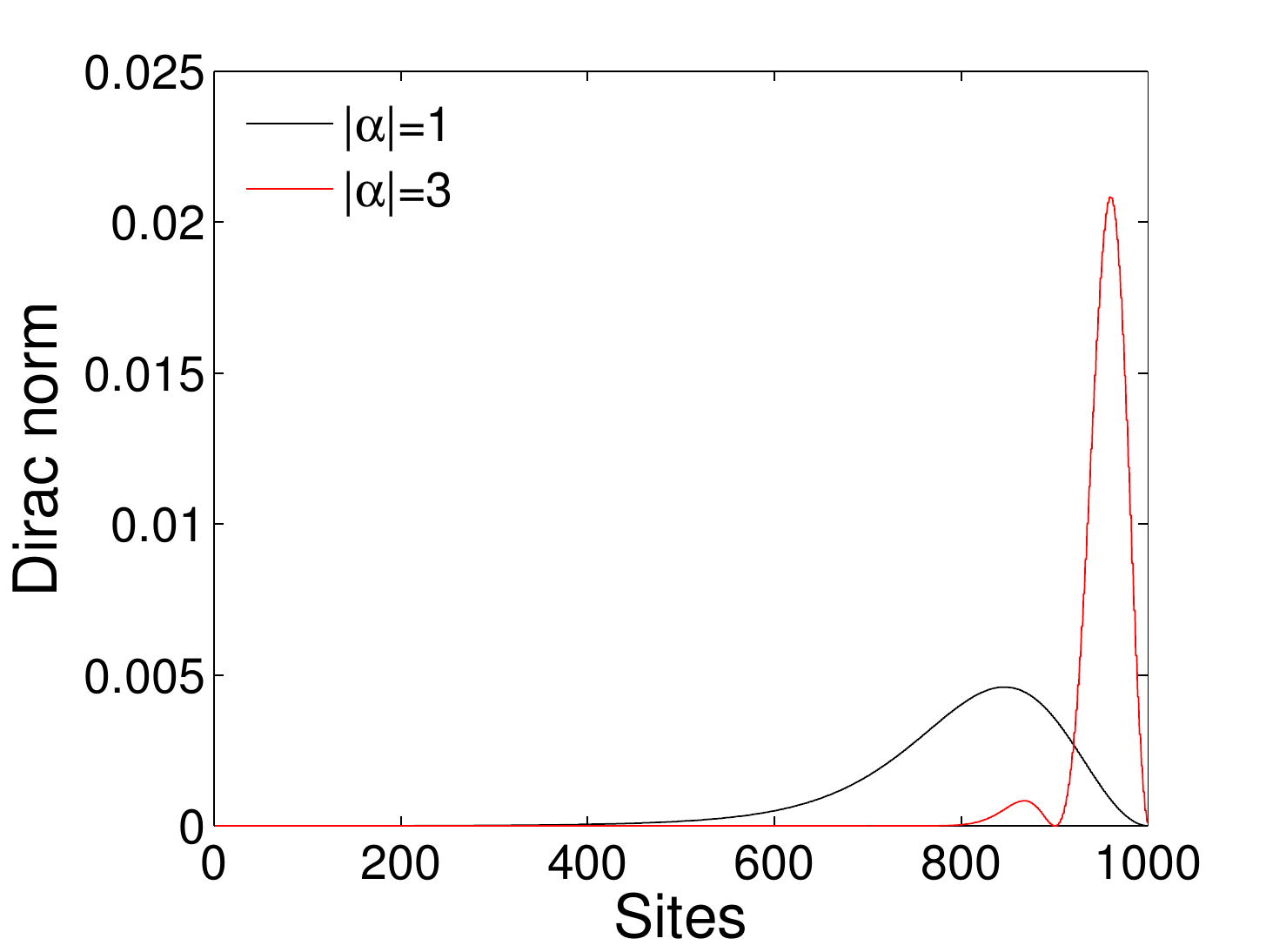}}		
\subfigure[] {\includegraphics[height=2in,width=2.2in]{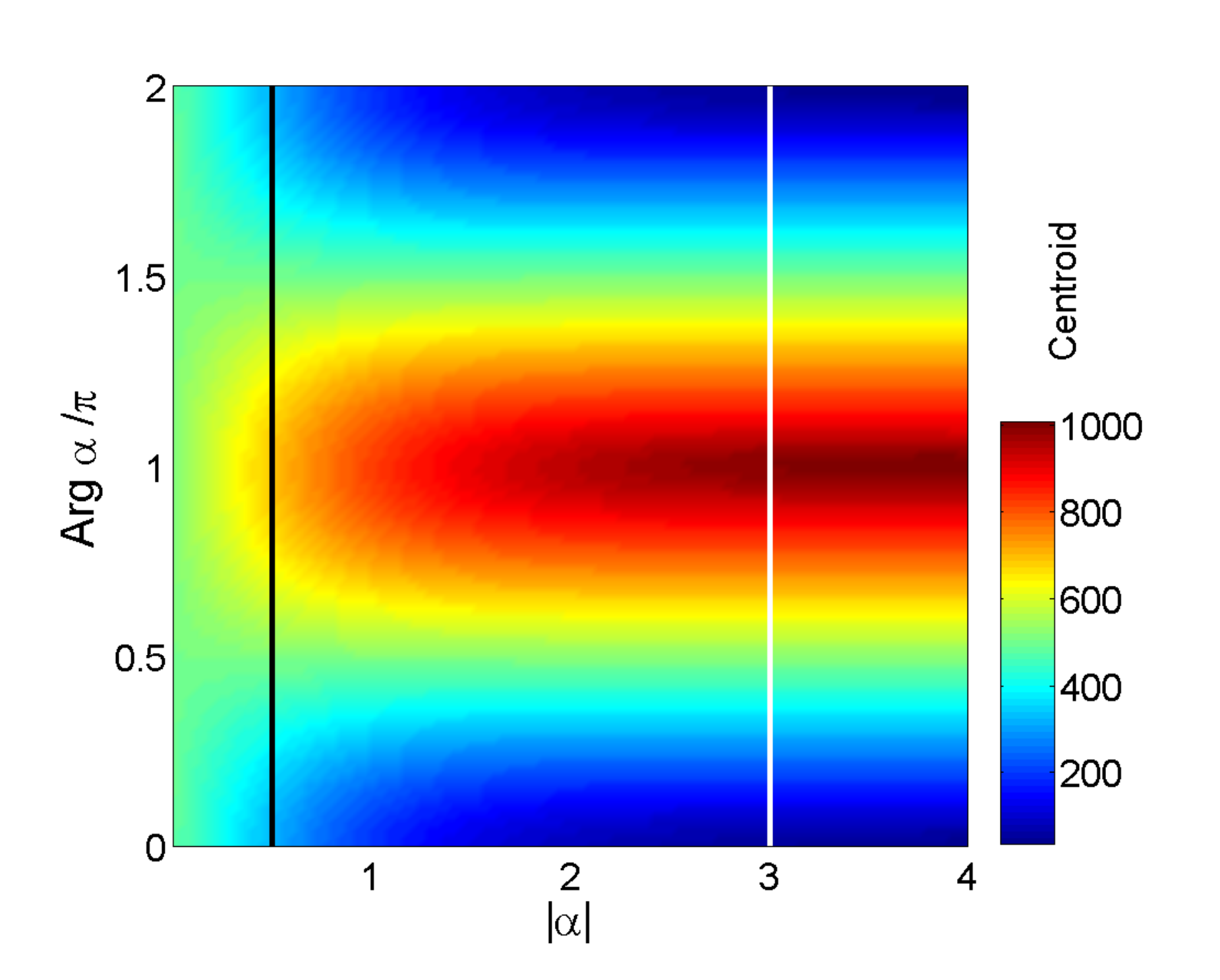}} %
\subfigure[] {\includegraphics[height=2in,width=2.2in]{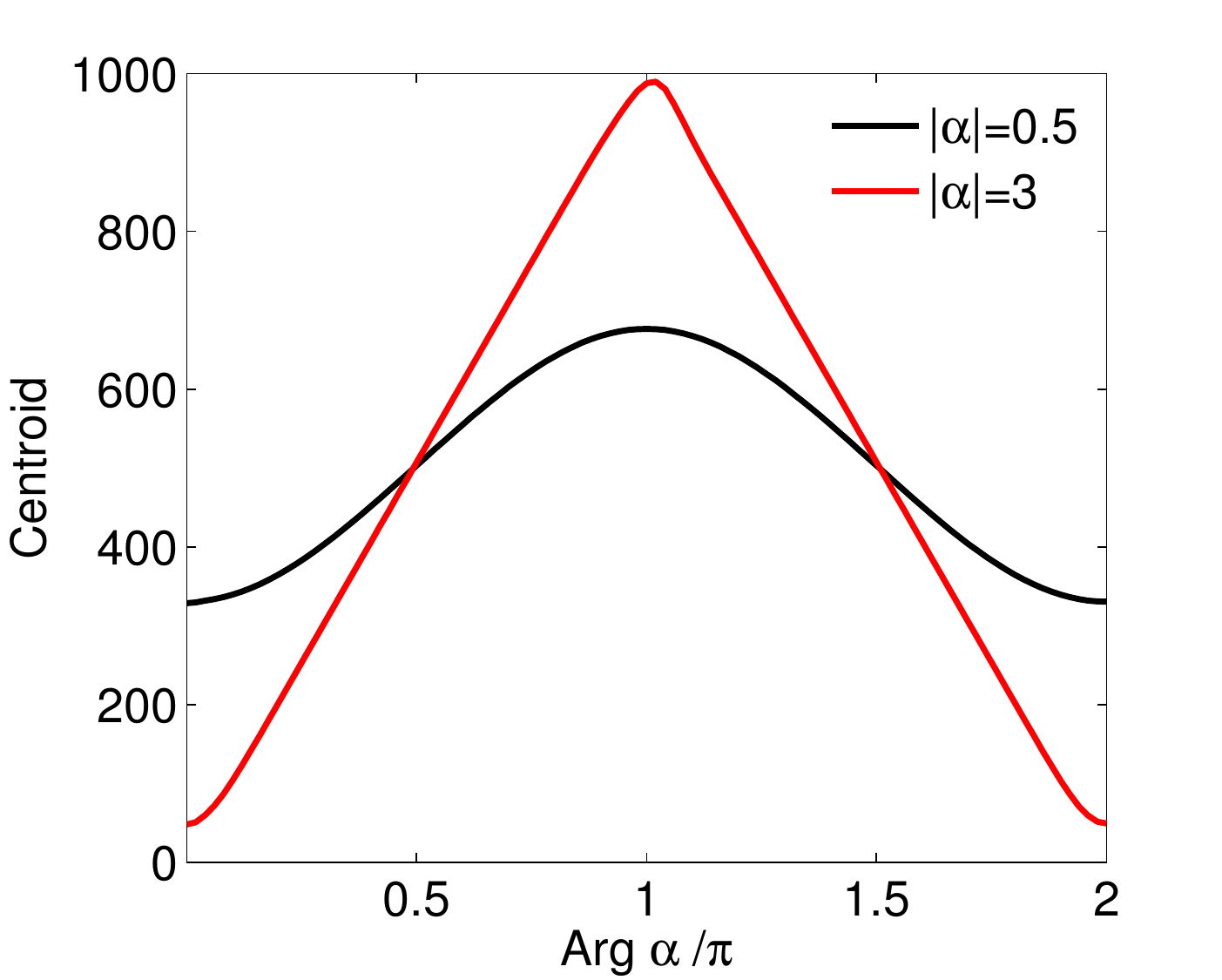}} %
\subfigure[] {\includegraphics[height=2in,width=2.2in]{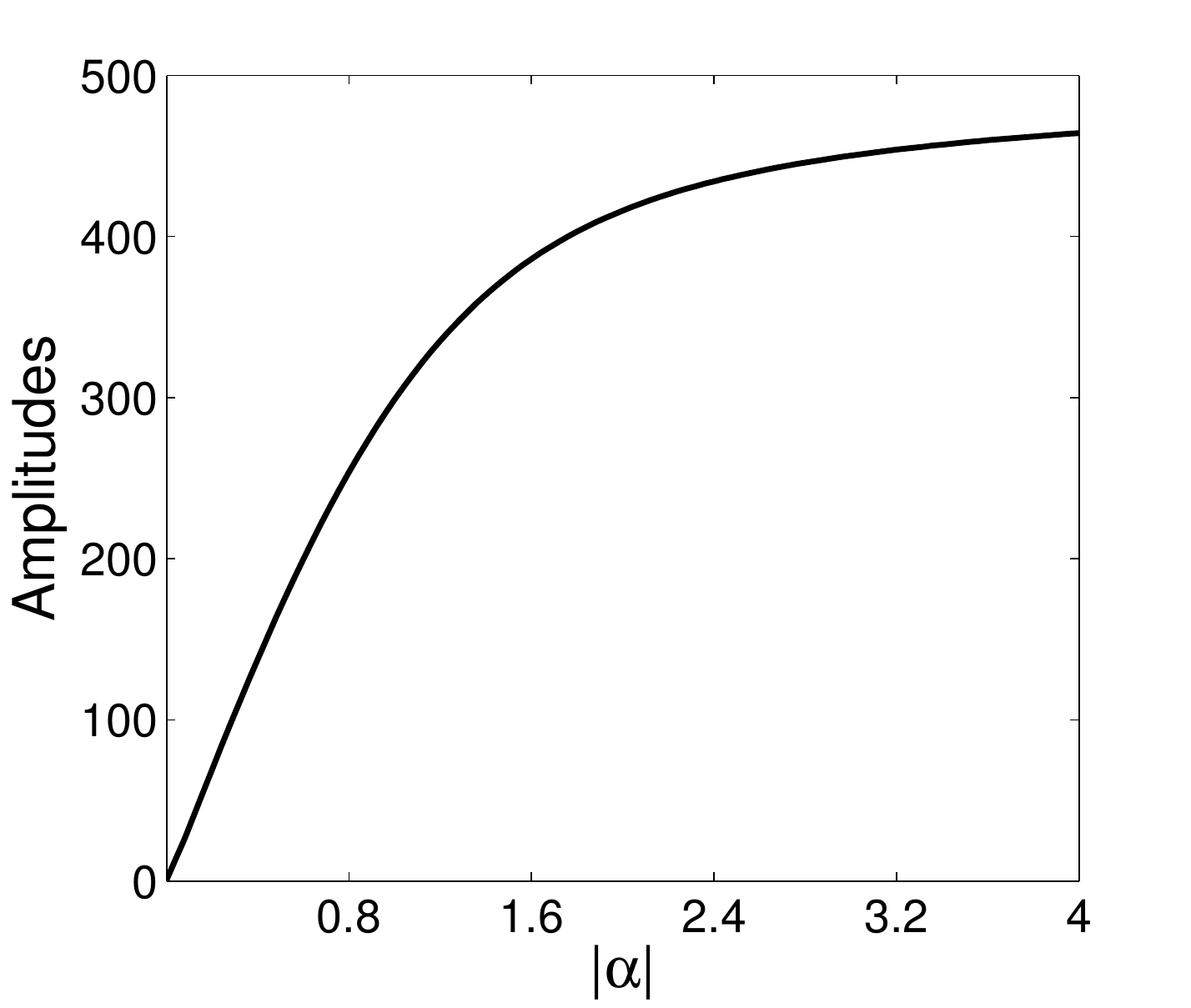}} 
\caption{(Color online) Plots of profiles of Dirac norm of the coherent
state from Eq. (\protect\ref{P_D}) with $\left\vert \protect\alpha %
\right\vert =0.1$ (black), $\left\vert \protect\alpha \right\vert =0.3$
(blue),\ and $\left\vert \protect\alpha \right\vert =0.5$ (red) with
different phases (a) 0, (b) $\protect\pi /2$ and (c) $\protect\pi$. (d)-(f) The same plot for the
cases with $\left\vert \protect\alpha \right\vert =1$ (black),\ and $%
\left\vert \protect\alpha \right\vert =3$ (red).
We can see that shape of the state for small $\left\vert \protect\alpha %
\right\vert $ is similar to the canonical coherent state, while deforms
strongly for large $\left\vert \protect\alpha \right\vert $. The position of
center of mass\ arises from the deformation of the wavepacket.\ (g) 3D plot of the
center of mass of the wavepackets of the states as the function of $\protect%
\alpha $ obtained from Eq. (\protect\ref{COM}). It is periodic function of the phase
of $\protect\alpha $.\ (h) The center of mass of the wavepackets
obtained from Eq. (\protect\ref{COM}) with $\left\vert \protect\alpha %
\right\vert =0.5$ (black)\ and $\left\vert \protect\alpha \right\vert =3$
(red), which correspond to the black and white lines in (g). It indicates that\
the periodic function is sinusoidal wave for small $\left\vert \protect%
\alpha \right\vert $ but tends to triangle wave as $\left\vert \protect%
\alpha \right\vert $\ increases. (i) Amplitudes of the center of mass as the
function of $|\protect\alpha |$. We see that it is linear for small $|%
\protect\alpha |$, which is similar to the canonical coherent state.\ The
parameters for the SSH chain are $N=500$, $\protect\delta =0.9$ and $\protect%
\gamma =1.8$. }
\label{fig2}
\end{figure*}

We consider a non-Hermitian SSH chain with staggered balanced gain and loss.
The simplest tight-binding model with these features is
\begin{eqnarray}
H &=&(1+\delta )\sum_{j=1}^{N}a_{j}^{\dag }b_{j}+(1-\delta
)\sum_{j=1}^{N-1}b_{j}^{\dag }a_{j+1}+\mathrm{H.c.}  \notag \\
&&+i\gamma \sum_{j=1}^{N}(a_{j}^{\dag }a_{j}-b_{j}^{\dag }b_{j}),  \label{H}
\end{eqnarray}%
where $\delta $ and $i\gamma $, are the distortion factor with unit
tunneling constant and the alternating imaginary potential magnitude,
respectively. Here $a_{l}^{\dag }$ and $b_{l}^{\dag }$ are the creation
operator of the particle at the $l$th site in $A$ and $B$\ sub-lattices. The
particle can be fermion or boson, depending on their own commutation
relations. A sketch of the lattice is shown in Fig.~\ref{Fig1}(a). In the
absence of the staggered potentials, the SSH model \cite{SSH} has served as
a paradigmatic example of the $1$-D system supporting topological character
\cite{Zak}. It has an extremely simple form but well manifests the typical
feature of topological insulating phase, and the transition between
non-trivial and trivial topological phases, associated with the number of
zero energy edge states as the topological invariant \cite{Asboth}. For
nonzero $\gamma $, it is still a $\mathcal{PT}$-symmetry. Here, the time
reversal operation $\mathcal{T}$ is such that $\mathcal{T}i\mathcal{T}=-i$,
while the effect of the parity is such that $\mathcal{P}a_{l}\mathcal{P}%
=b_{N+1-l}$ and $\mathcal{P}b_{l}\mathcal{P}=a_{N+1-l}$. Applying operators $%
\mathcal{P}$ and $\mathcal{T}$ on the Hamiltonian (\ref{H}), one has $\left[
\mathcal{T},H\right] \neq 0$ and $\left[ \mathcal{P},H\right] \neq 0$, but $%
\left[ \mathcal{PT},H\right] =0$. According to the non-Hermitian quantum
theory, such a Hamiltonian may have fully real spectrum within a certain
parameter region. The boundary of the region is the critical point of
quantum phase transition associated with $\mathcal{PT}$-symmetry breaking.
The system with periodic boundary condition has been studied in Ref. \cite%
{HWH}. We will show that open boundary condition leads to different dynamical
behavior.

According to the Appendix, in the strong dimerization limit $1+\delta \gg
1-\delta $, the Hamiltonian can be diagonalized as the form%
\begin{equation}
H=\sum_{k}\varepsilon _{k}(\overline{\alpha }_{k}\alpha _{k}-\overline{\beta
}_{k}\beta _{k}),  \label{H_k}
\end{equation}%
where the operators are%
\begin{eqnarray}
\alpha _{k} &=&\sqrt{\frac{2}{N+1}}\sum_{j=1}^{N}(-1)^{j}\sin \left(
kj\right) \frac{a_{j}+e^{-i\varphi _{k}}b_{j}}{1+ie^{-i\varphi _{k}}}, \\
\beta _{k} &=&\sqrt{\frac{2}{N+1}}\sum_{j=1}^{N}(-1)^{j}\sin \left(
kj\right) \frac{a_{j}-e^{i\varphi _{k}}b_{j}}{1+ie^{-i\varphi _{k}}},
\end{eqnarray}%
and their counterparts are%
\begin{eqnarray}
\overline{\alpha }_{k} &=&\sqrt{\frac{2}{N+1}}\sum_{j=1}^{N}(-1)^{j}\sin
\left( kj\right) \frac{a_{j}^{\dag }+e^{-i\varphi _{k}}b_{j}^{\dag }}{%
1-ie^{-i\varphi _{k}}}, \\
\overline{\beta }_{k} &=&\sqrt{\frac{2}{N+1}}\sum_{j=1}^{N}(-1)^{j}\sin
\left( kj\right) \frac{a_{j}^{\dag }-e^{i\varphi _{k}}b_{j}^{\dag }}{%
1-ie^{-i\varphi _{k}}}.
\end{eqnarray}%
Here the dispersion relation and phase are%
\begin{eqnarray}
\varepsilon _{k} &=&\sqrt{[(1+\delta )-(1-\delta )\cos k]^{2}-\gamma ^{2}},
\\
\tan \varphi _{k} &=&\frac{\gamma }{\varepsilon _{k}},
\end{eqnarray}%
with $k=\frac{(n+1)\pi }{N+1},$ $n=0,1,2,3,...,N-1$.

The non-Hermitian operator $H$ in Eq.~(\ref{H_k}) is in diagonal form, since
$\alpha _{k}$, $\overline{\alpha }_{k}$, $\beta _{k}$, and $\overline{\beta }%
_{k}$ are canonical conjugate operators, obeying the canonical commutation
relations
\begin{eqnarray}
\left[ \alpha _{k},\overline{\alpha }_{k^{\prime }}\right] _{\pm } &=&\left[
\beta _{k},\overline{\beta }_{k^{\prime }}\right] _{\pm }=\delta
_{kk^{\prime }},  \notag \\
\left[ \alpha _{k},\alpha _{k^{\prime }}\right] _{\pm } &=&\left[ \beta
_{k},\beta _{k^{\prime }}\right] _{\pm }=0,  \notag \\
\left[ \overline{\alpha }_{k},\overline{\alpha }_{k^{\prime }}\right] _{\pm
} &=&\left[ \overline{\beta }_{k},\overline{\beta }_{k^{\prime }}\right]
_{\pm }=0,  \notag \\
\left[ \alpha _{k},\overline{\beta }_{k^{\prime }}\right] _{\pm } &=&\left[
\overline{\alpha }_{k},\overline{\beta }_{k^{\prime }}\right] _{\pm }=0,
\notag \\
\left[ \alpha _{k},\beta _{k^{\prime }}\right] _{\pm } &=&\left[ \overline{%
\alpha }_{k},\beta _{k^{\prime }}\right] _{\pm }=0.  \label{canon_CR}
\end{eqnarray}%
The system is pseudo-Hermitian since it can either have fully real spectrum
or complex spectrum with complex conjugation pair imaginary levels. We
emphasize that the canonical conjugate pairs appearing in Eqs. (\ref%
{canon_CR}) are not simply defined by the Hermitian conjugate operation,
i.e. $\overline{\alpha }_{k}\neq \alpha _{k}^{\dagger }$ and $\overline{%
\beta }_{k}\neq \beta _{k}^{\dagger }$, which is differ from that in a
Hermitian regime. Although the Eqs. (\ref{canon_CR}) is obtained by the
approximation in the strong dimerization limit, it should be hold within all
the range of parameter when the exact expression of operators $(\overline{%
\alpha }_{k},\alpha _{k},\overline{\beta }_{k},\beta _{k})$\ is applied.

We note that the spectrum $\varepsilon _{k}$ consists of two branches
separated by an energy gap $\Delta =\sqrt{4\delta ^{2}-\gamma ^{2}}$.
Obviously, it displays a full real spectrum within the region of $4\delta
^{2}\geq \gamma ^{2}$, which is refer to as unbroken $\mathcal{PT}$-symmetry
region. Beyond this region, the imaginary eigenvalues appears and the $%
\mathcal{PT}$ symmetry of the corresponding eigenfunction is broken
simultaneously according to the non-Hermitian quantum theory. Notice that,
when the onset of the $\mathcal{PT}$ symmetry breaking begins, the band gap
vanishes. We are interested in the system at the critical point, at which
the gap close, separating two quantum phases, with full real and complex
spectra, respectively. The system reaches the critical point when $\gamma $\
takes\ $\gamma _{c}=2\delta $. The zero modes corresponds to $k_{c}=\frac{%
\pi }{N+1}$, which results in $\tan \varphi _{k_{c}}\rightarrow \infty $ and%
\begin{eqnarray}
\alpha _{k_{c}} &=&\beta _{k_{c}}=\sqrt{\frac{2}{N+1}}\sum_{j=1}^{N}(-1)^{j}%
\sin \left( k_{c}j\right) \frac{a_{j}-ib_{j}}{2},  \notag \\
\overline{\alpha }_{k_{c}} &=&\overline{\beta }_{k_{c}}=\sqrt{\frac{2}{N+1}}%
\sum_{j=1}^{N}(-1)^{j}\sin \left( k_{c}j\right) \frac{a_{j}^{\dag
}-ib_{j}^{\dag }}{1-ie^{-i\varphi _{k_{c}}}},  \notag \\
&&\left\langle 0\right\vert \alpha _{k_{c}}\overline{\alpha }%
_{k_{c}}\left\vert 0\right\rangle =\left\langle 0\right\vert \beta _{k_{c}}%
\overline{\beta }_{k_{c}}\left\vert 0\right\rangle =0,
\end{eqnarray}%
i.e., two zero modes become identical, where $\left\vert 0\right\rangle $\
is vacuum state of particle operators $\left( a_{j},b_{j}\right) $. In
contrast to a Hermitian system, two zero-energy eigen states coalescence at
the EP point, rather than degeneracy point. At this situation, the spectrum
around zero energy can be expressed as%
\begin{equation}
\varepsilon _{k}\approx \sqrt{2\delta (1-\delta )}k,
\end{equation}%
which is linear in $k$. The spectra of the lattice is shown in Fig.\ \ref%
{Fig1}(a).

\section{Hermitian dynamics in sub-space}

\label{Hermitian dynamics in sub-space}

In this section, we investigate the dynamics in the system within unbroken
region. For simplicity, we only focus on single-particle invariant space.
The obtained result can be extended to the many-particle invariant subspace
due to the canonical commutation relations in Eqs.~(\ref{canon_CR}).
According to the non-Hermitian quantum theory, the eigenstates of a
pseudo-Hermitian Hamiltonian can construct a set of biorthogonal bases in
association with the eigenstates of its Hermitian conjugate. Similar to the
works \cite{LJin84}\cite{HWH}, eigenstates $\{\overline{\alpha }%
_{k}\left\vert 0\right\rangle ,\overline{\beta }_{k}\left\vert
0\right\rangle \}$ of $H$ and eigenstates $\{\alpha _{k}^{\dagger
}\left\vert 0\right\rangle ,\beta _{k}^{\dagger }\left\vert 0\right\rangle \}
$ of $H^{\dagger }$ are the biorthogonal bases of the single-particle
invariant subspace. In general, the eigenstates of a non-Hermitian
Hamiltonian are not orthogonal under the Dirac inner product due to the
non-Hermiticity of the Hamiltonian. However, we note that orthogonality
between the eigenstates with different $k$ in the Dirac inner product still
maintain due to following quasi-canonical commutation relations
\begin{eqnarray}
\left[ \alpha _{k},\alpha _{k^{\prime }}^{\dagger }\right] _{\pm } &=&\left[
\beta _{k},\beta _{k^{\prime }}^{\dagger }\right] _{\pm }=\frac{1}{1+\sin
\varphi _{k}}\delta _{kk^{\prime }},  \notag \\
\left[ \overline{\alpha }_{k}^{\dagger },\overline{\alpha }_{k^{\prime }}%
\right] _{\pm } &=&[\overline{\beta }_{k}^{\dagger },\overline{\beta }%
_{k^{\prime }}]_{\pm }=\frac{1}{1-\sin \varphi _{k}}\delta _{kk^{\prime }},
\notag \\
\left[ \beta _{k},\alpha _{k^{\prime }}^{\dagger }\right] _{\pm } &=&\frac{%
1-e^{i2\varphi _{k}}}{2\left( 1+\sin \varphi _{k}\right) }\delta
_{kk^{\prime }}, \\
\left[ \overline{\alpha }_{k}^{\dagger },\overline{\beta }_{k^{\prime }}%
\right] _{\pm } &=&\frac{1-e^{i2\varphi _{k}}}{2\left( 1-\sin \varphi
_{k}\right) }\delta _{kk^{\prime }},  \notag \\
\left[ \alpha _{k},\overline{\alpha }_{k^{\prime }}^{\dagger }\right] _{\pm
} &=&[\beta _{k},\overline{\beta }_{k^{\prime }}^{\dagger }]_{\pm }=0,
\notag \\
\lbrack \alpha _{k},\overline{\beta }_{k^{\prime }}^{\dagger }]_{\pm } &=&%
\left[ \beta _{k},\overline{\alpha }_{k^{\prime }}^{\dagger }\right] _{\pm
}=0.  \notag
\end{eqnarray}%
We use the term quasi\ due to the absence of orthogonality between the
eigenmodes of $\overline{\alpha }_{k}$ and $\overline{\beta }_{k}$. We note
that if one only consider the particle with positive energy (particle) or
particle with negative energy (hole), the non-Hermitian system appears as a
Hermitian one due to the commutation relations\ for two particles (or
holes). This is crucial for this work.

\section{Simple harmonic oscillation}

\label{Simple harmonic oscillation}

\begin{figure*}[!t]
\centering
\subfigure[$\ \protect\alpha
=0.5$]{\includegraphics[height=1.89in,width=2.32in]{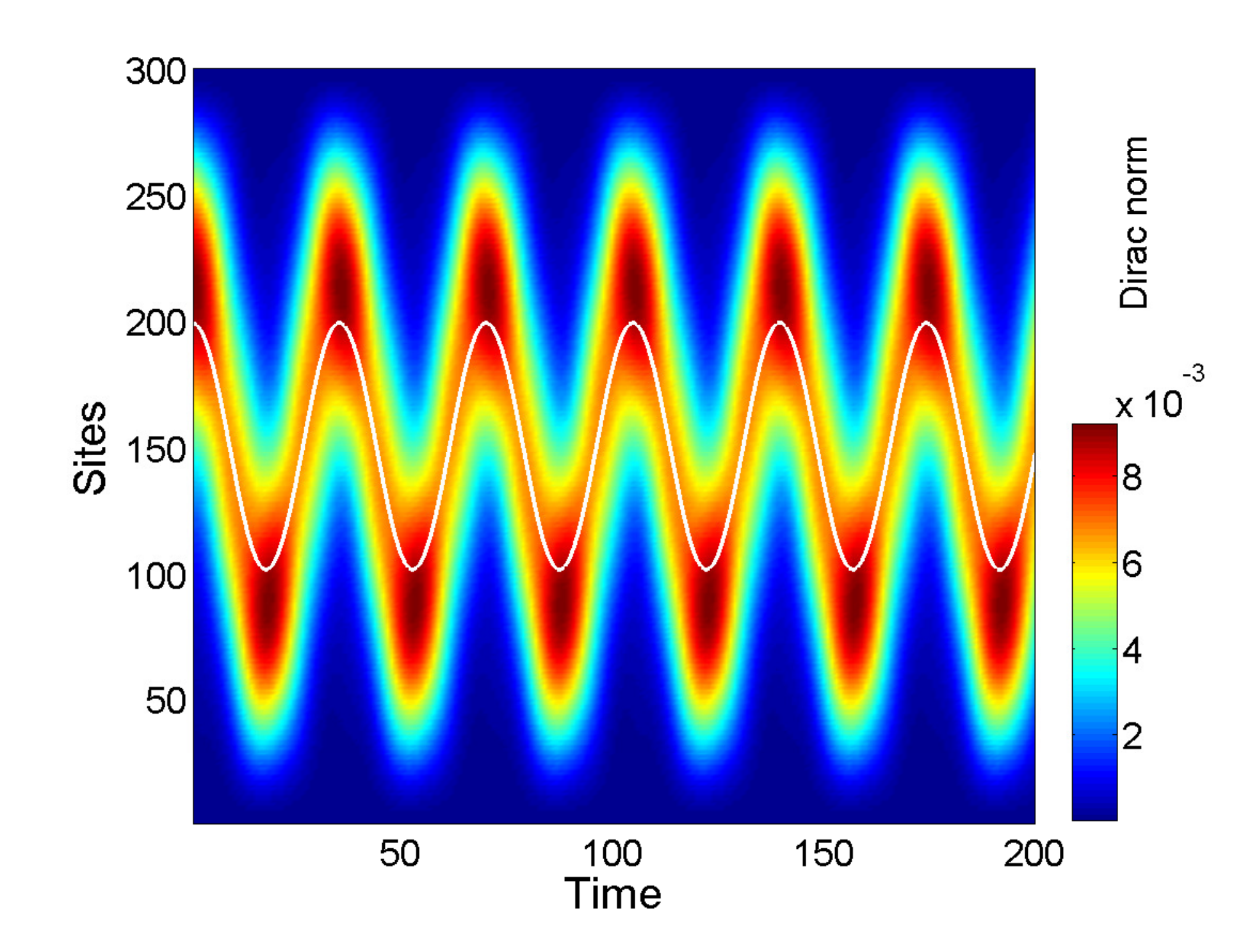}}
\subfigure[$\ \protect\alpha
=1$]{\includegraphics[height=2in,width=2.32in]{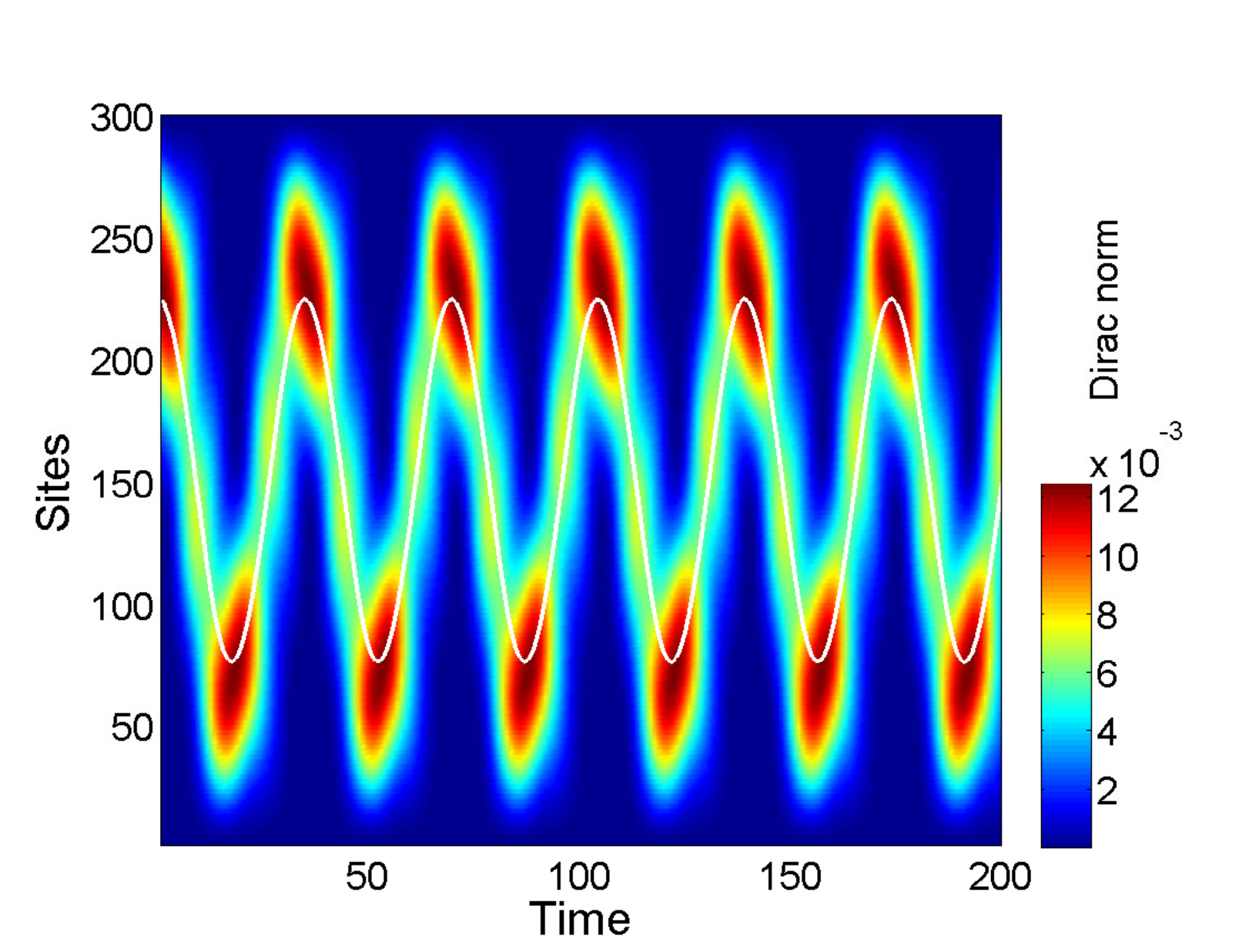}}
\subfigure[$\
\protect\alpha =3$]{\includegraphics[height=2in,width=2.32in]{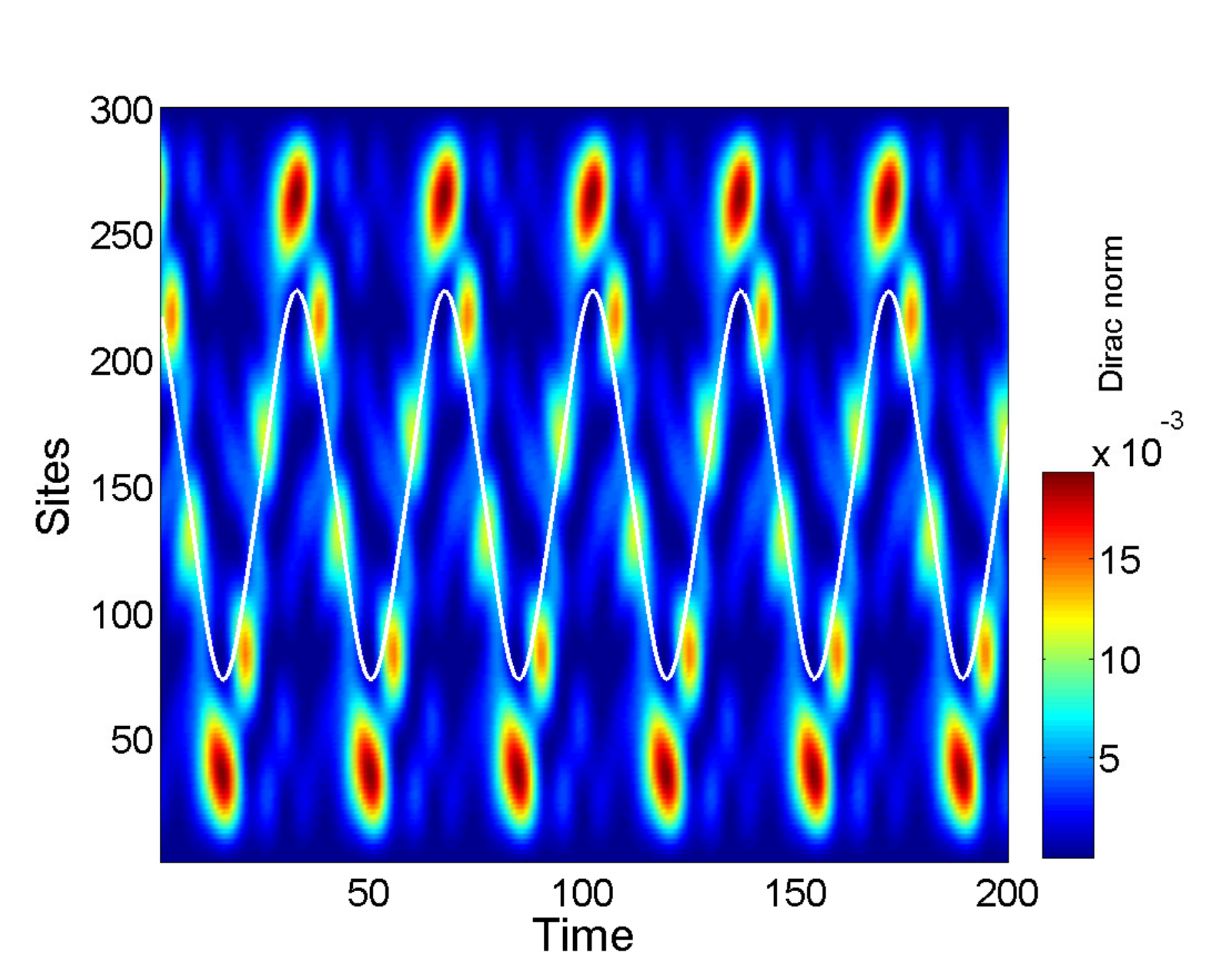}}
\subfigure[$\ \protect\alpha
=4$]{\includegraphics[height=2in,width=2.34in]{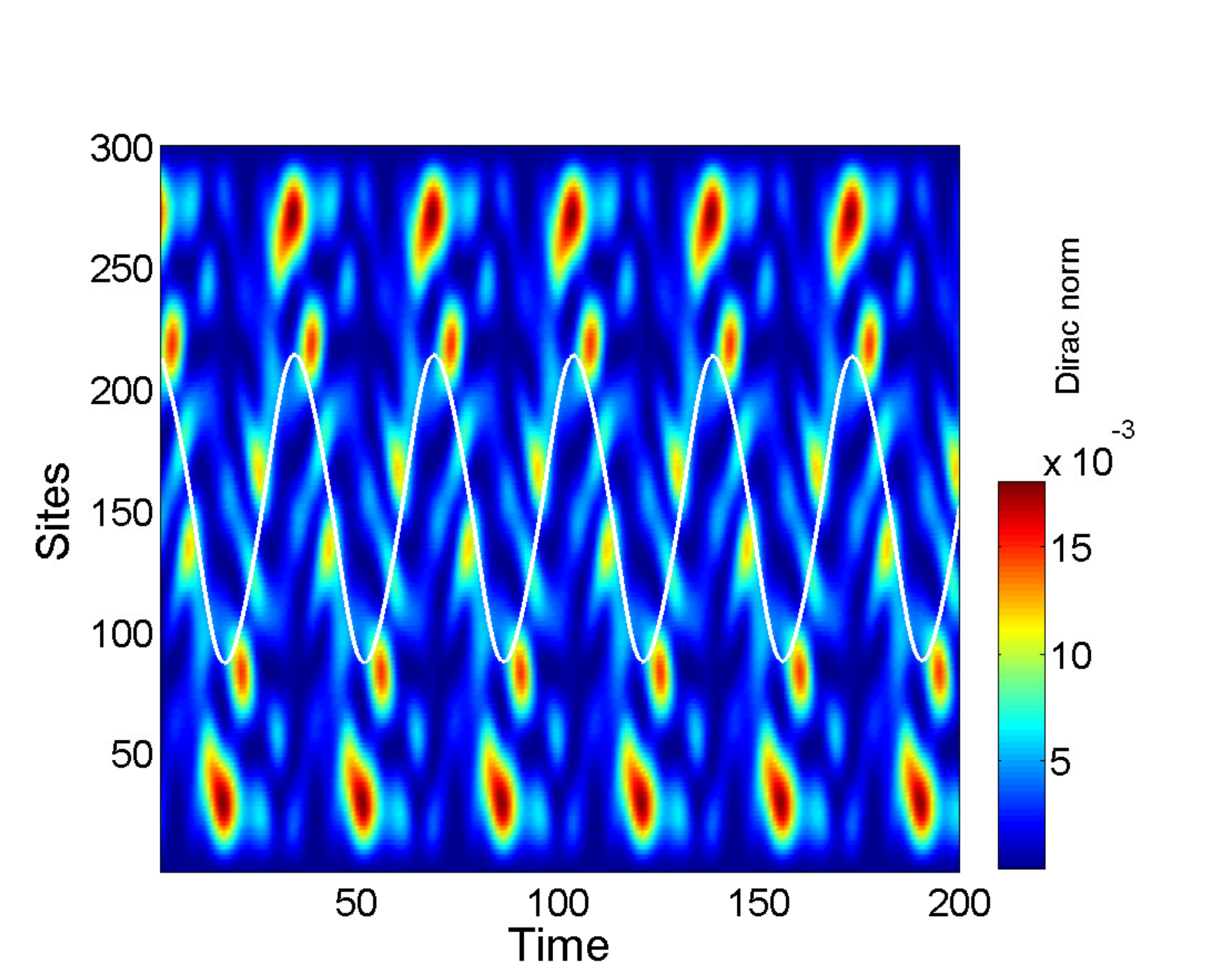}}
\subfigure[$\
\protect\alpha =6$]{\includegraphics[height=2in,width=2.32in]{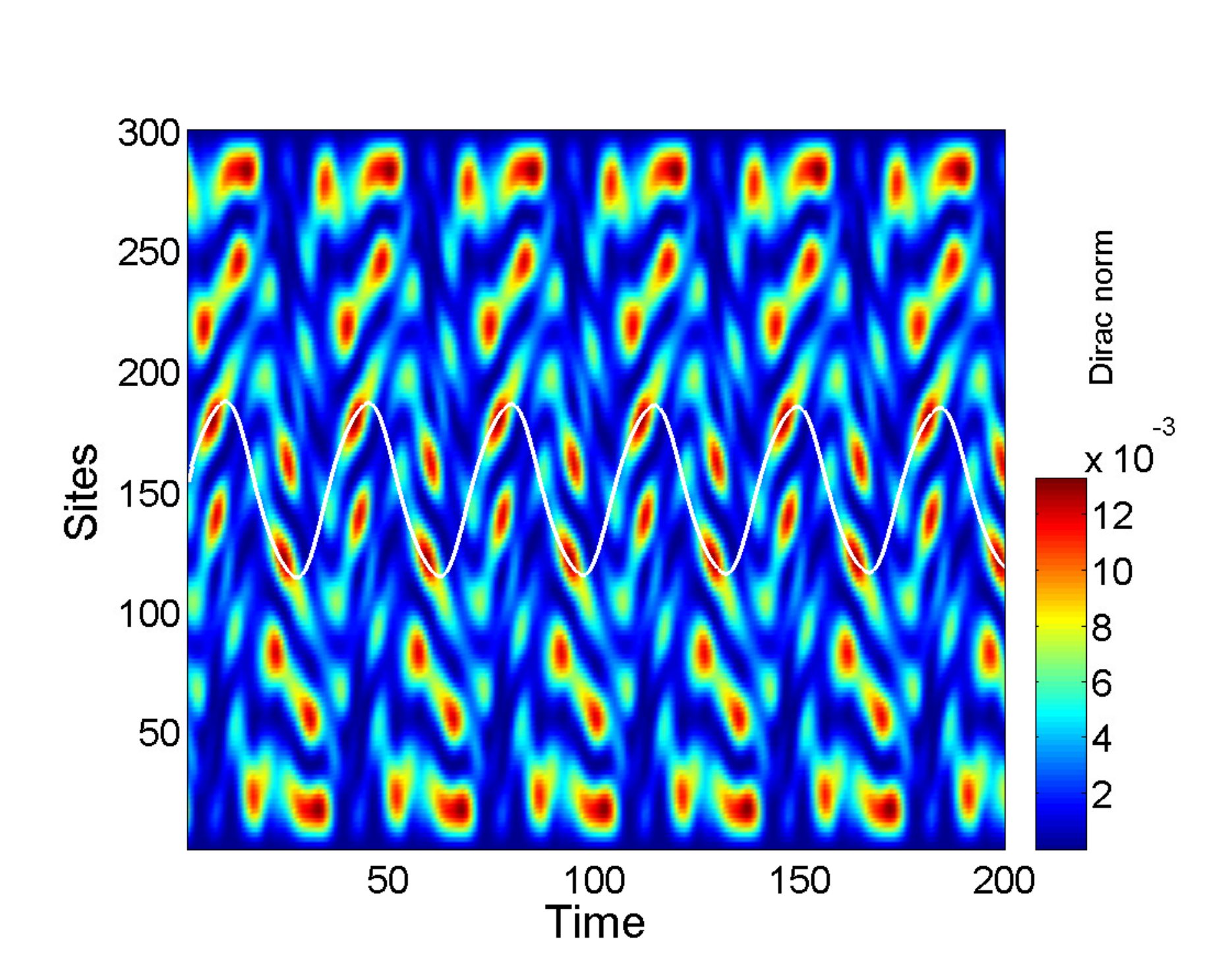}}
\subfigure[$\ \protect\alpha
=12$]{\includegraphics[height=2in,width=2.32in]{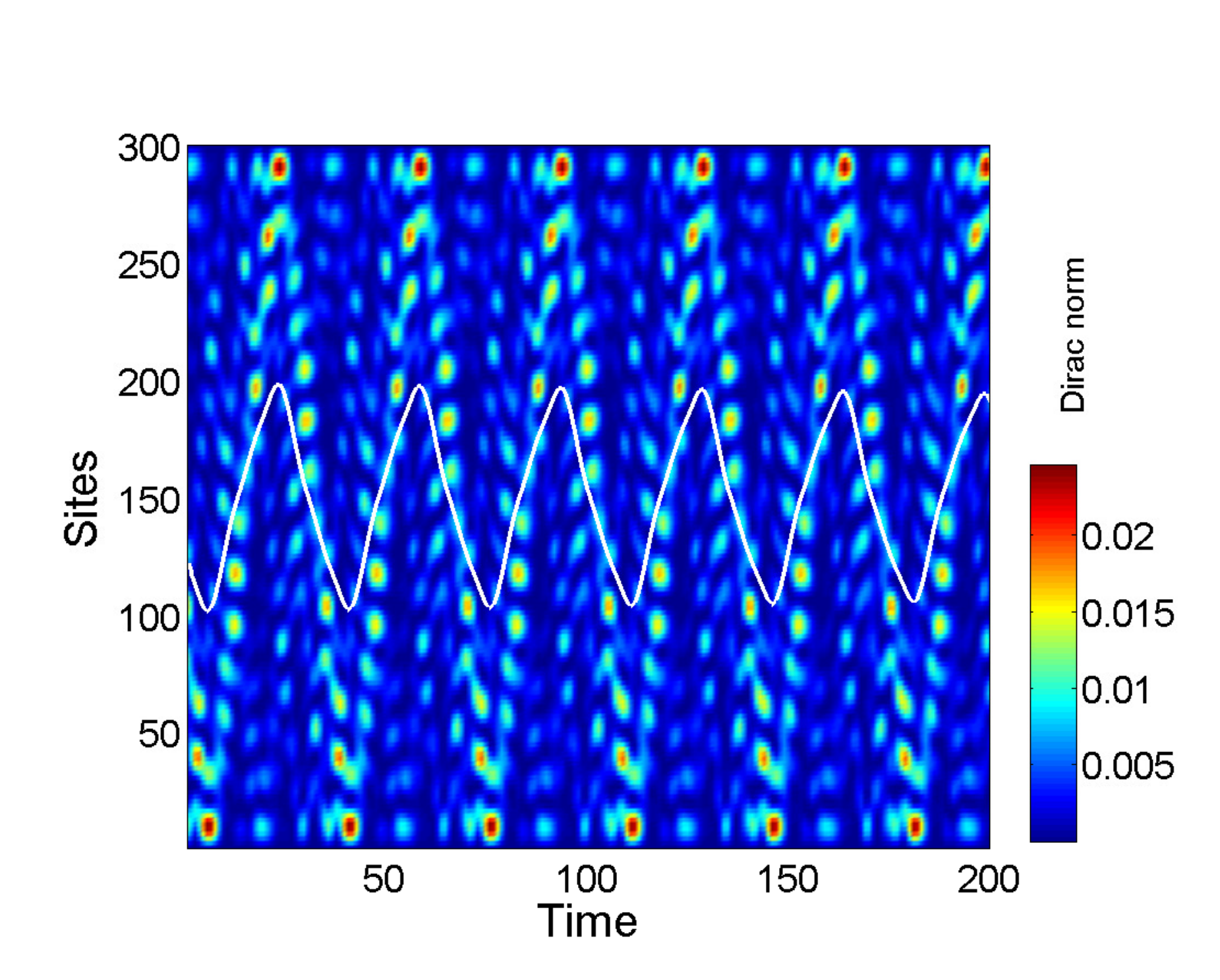}}
\caption{(Color online) Plots of numerical simulations\ for the profiles of
Dirac norm of the time evolution for initial coherent states with several
typical $\protect\alpha $ for the SSH chain with parameters $N=150$, $%
\protect\delta =0.9$ and $\protect\gamma =1.8$. (a) $\protect\alpha =0.5$.
(b) $\protect\alpha =1$, (c) $\protect\alpha =3$, (d) $\protect\alpha =4$,
(e) $\protect\alpha =6$, and (f) $\protect\alpha =12$, respectively. The
white solid lines indicate the center of mass of the wavepackets. It shows
that the trajectory is sinusoidal wave for small $\protect\alpha $, while
close to triangle wave for large $\protect\alpha $. Here the time is in
units of $\ 20{J}^{-1}$, where $J$ is the scale of the Hamiltonian and we
take $J=1$.}
\label{fig3}
\end{figure*}

In this section, we investigate the dynamics in the gapless system. The
single-particle eigen vectors can be expressed as%
\begin{equation}
\left\vert \psi _{n}^{\pm }\right\rangle =\frac{1}{\sqrt{N+1}}%
\sum_{j=1}^{N}(-1)^{j}\sin \left( kj\right) \left( a_{j}^{\dag }\pm e^{\mp
i\varphi _{k}}b_{j}^{\dag }\right) \left\vert 0\right\rangle ,
\end{equation}%
\ where $k$ is redefined as%
\begin{equation}
k=\frac{(n+1)\pi }{N+1},\text{ }n\in \lbrack 0,N-1].
\end{equation}%
The corresponding eigen energy is%
\begin{equation}
E_{n}^{\pm }=\pm (n+1)\omega ,\text{ }\omega =\frac{\sqrt{2\delta (1-\delta )%
}\pi }{N+1},
\end{equation}%
approximately for not large $n$. Here $\left\vert \psi _{n}^{\pm
}\right\rangle $\ is normalized in the framework of Dirac inner product,
i.e., $\langle \psi _{m}^{\pm }\left\vert \psi _{n}^{\pm }\right\rangle
=\delta _{mn}$. We note that $\left\vert \psi _{0}^{+}\right\rangle
=\left\vert \psi _{0}^{-}\right\rangle $ is the coalescing eigenstates and $%
\langle \psi _{n}^{+}\left\vert \psi _{n}^{-}\right\rangle \neq 0$. When we
focus on the positive-energy region, it is similar to the light field mode
in a rectangular cavity, standing wave with uniform energy-level spacing. In
the rest of paper, we concentrate on the positive energy Hilbert space and
denote $\left\vert \psi _{n}\right\rangle =\left\vert \psi
_{n}^{+}\right\rangle $.

Following the idea for a quantum simple harmonic oscillator system, we
construct a coherent-like state

\begin{equation}
\left\vert \alpha \right\rangle =e^{-\frac{\left\vert \alpha \right\vert ^{2}%
}{2}}\sum_{n=0}\frac{\alpha ^{n}}{\sqrt{n!}}\left\vert \psi
_{n}\right\rangle .
\end{equation}%
We define the number operator $\hat{n}$ by%
\begin{equation}
\hat{n}\left\vert \psi _{n}\right\rangle =n\left\vert \psi _{n}\right\rangle
,
\end{equation}%
and find that%
\begin{equation}
\bar{n}=\left\langle \alpha \right\vert \hat{n}\left\vert \alpha
\right\rangle =\left\vert \alpha \right\vert ^{2},
\end{equation}%
which indicates the significance of $\alpha $. The position state $%
\left\vert l\right\rangle $ with $l\in \lbrack 1,2N]$\ is%
\begin{equation*}
\left\vert 2j-1\right\rangle =a_{j}^{\dag }\left\vert 0\right\rangle ,\text{
}\left\vert 2j\right\rangle =b_{j}^{\dag }\left\vert 0\right\rangle ,
\end{equation*}%
with $j\in \lbrack 1,N]$. The profile of distribution of Dirac probability
of the state can be obtained as
\begin{equation}
P_{\mathrm{D}}\left( l\right) =\left\vert \langle l\left\vert \alpha
\right\rangle \right\vert ^{2},  \label{P_D}
\end{equation}%
and the center of mass of the wavepacket is\
\begin{equation}
r_{c}=\sum_{l=1}^{2N}l\left\vert \langle l\left\vert \alpha \right\rangle
\right\vert ^{2}.  \label{COM}
\end{equation}%
To demonstrate the property of coherent-like state, we plot Eqs. (\ref{P_D})
and (\ref{COM}) for some typical $\alpha $\ in Fig. \ref{fig2}. It shows
that $r_{c}$ is determined by the phase $\arg (\alpha )$ and the amplitude $%
\left\vert \alpha \right\vert $ of $\alpha $. The detailed features of the
state can be understood by the following anaysis since the phase of $\alpha $%
\ is equivalent to the phase factor arising from the time evolution.

Now we turn to the dynamics of the state $\left\vert \alpha \right\rangle $.
Under the strong dimerization approximation, the evolved state is

\begin{eqnarray}
U(t)\left\vert \alpha \right\rangle  &=&e^{-\frac{\left\vert \alpha
\right\vert ^{2}}{2}}\sum_{n=0}\frac{\alpha ^{n}}{\sqrt{n!}}e^{-i(n+1)\omega
t}\left\vert n\right\rangle   \notag \\
&=&e^{-\frac{\left\vert \alpha \right\vert ^{2}}{2}-i\omega t}\sum_{n=0}%
\frac{\left( \alpha e^{-i\omega t}\right) ^{n}}{\sqrt{n!}}\left\vert
n\right\rangle ,
\end{eqnarray}%
which shows that factor $\omega t$\ takes the role of the phase of $\alpha $%
. Then the trajectory of the centroid%
\begin{equation}
r_{c}(t)=\sum_{l=1}^{2N}l\left\vert \left\langle l\right\vert U(t)\left\vert
\alpha \right\rangle \right\vert ^{2},
\end{equation}%
can be estimated directly from above analysis. In the Appendix, we get the
following results: (i) In the small $\left\vert \alpha \right\vert $\ limit,
we have%
\begin{equation}
r_{c}(t)\approx -\frac{64N}{9\pi ^{2}}\left\vert \alpha \right\vert \cos
(\omega t-\arg \alpha )+N,
\end{equation}%
which is a sinusoidal wave. (ii) In the $\left\vert \alpha \right\vert \gg 1$%
\ limit, we have%
\begin{equation}
r_{c}(t)=2N\left( \frac{\omega t}{\pi }-2n\right) \times \left\{
\begin{array}{c}
1,\text{ }t\in \lbrack 0,\frac{\pi }{\omega })+\frac{2n\pi }{\omega } \\
-1,\text{ }t\in \lbrack -\frac{\pi }{\omega },0)+\frac{2n\pi }{\omega }%
\end{array}%
,\text{ }n\in Z\right. ,
\end{equation}%
which is a triangle wave. It accords with the plots in Fig. \ref{fig2}.

All the above analysis are based on the assumption of the strong
dimerization limit. We are interested in the case with moderate value of $%
\delta $. To verify our result and demonstrate the extent to which the
harmonic oscillation also exists, we perform numerical by exact
diagonalization for finite-size SSH chain with different $\alpha $. Plots in
Fig. \ref{fig3} is the profile of Dirac probability distribution of evolved
state for several typical cases. It shows that the trajectory is sinusoidal
wave for small $\alpha $, while close to triangle wave for large $\alpha $.
Surprisingly, the trajectory is always smooth even the wavepacket spreads
out in the chain.

\section{Summary}

\label{Summary}

In summary, we studied the dynamics of non-Hermitian SSH chain at EP. We
have shown that a pre-engineered state can exhibit perfect SHM. The underly
mechanism of such dynamic behavior are based on two conditions (EHSM): (i)
linear dispersion relation and (ii) long-wave length standing-wave modes.
Such two features even appear in a simple uniform chain ($H$ with $\delta
=\gamma =0$) but not coexist in a same set of eigenstates. It is hard to
design a Hermitian tight-binding chain satisfying such two conditions. This
fact highlights the advantage of the non-Hermitian system. In the present
system, the balance of distortion and staggered imaginary potentials leads
to a set of eigenstates possessing such two features. It is also associated
with the concept of quasi-Hermitian sub-space. If one only consider the
particle with positive energy (particle) or particle with negative energy
(hole), the non-Hermitian system appears as a Hermitian one, preserving the
Dirac probability. This feature also appears in a ring system \cite{HWH}, in
which a wavepacket moves periodically but with constant speed. The peculiar
dynamical behavior is a demonstration of rich potential resource of the
non-Hermitian system at\ EP. Our result indicates that novel Hermitian
dynamics can be realized by a non-Hermitian system.

\section{Appendix}

\subsection{Approximate solutions}

We start with the Hermitian Hamiltonian

\begin{equation}
H_{0}=(1+\delta )\sum_{j=1}^{N}a_{j}^{\dag }b_{j}+(1-\delta
)\sum_{j=1}^{N-1}b_{j}^{\dag }a_{j+1}+\mathrm{H.c..}
\end{equation}%
Introducing particle operators%
\begin{equation}
\alpha _{j}^{\dag }=\frac{1}{\sqrt{2}}(a_{j}^{\dag }+b_{j}^{\dag }),\beta
_{j}^{\dag }=\frac{1}{\sqrt{2}}(a_{j}^{\dag }-b_{j}^{\dag }),
\end{equation}%
or inversely%
\begin{equation}
a_{j}^{\dag }=\frac{1}{\sqrt{2}}(\alpha _{j}^{\dag }+\beta _{j}^{\dag
}),b_{j}^{\dag }=\frac{1}{\sqrt{2}}(\alpha _{j}^{\dag }-\beta _{j}^{\dag }),
\end{equation}%
we have

\begin{eqnarray}
a_{j}^{\dag }b_{j}+b_{j}^{\dag }a_{j} &=&\alpha _{j}^{\dag }\alpha
_{j}-\beta _{j}^{\dag }\beta _{j}, \\
b_{j}^{\dag }a_{j+1}+a_{j+1}^{\dag }b_{j} &=&\frac{1}{2}\alpha _{j}^{\dag
}\alpha _{j+1}-\frac{1}{2}\beta _{j}^{\dag }\beta _{j+1}  \notag \\
&&+\frac{1}{2}\alpha _{j}^{\dag }\beta _{j+1}-\frac{1}{2}\beta _{j}^{\dag
}\alpha _{j+1}+\mathrm{H.c..}
\end{eqnarray}%
Under the condition%
\begin{equation}
1+\delta \gg 1-\delta ,
\end{equation}%
we have%
\begin{equation}
b_{j}^{\dag }a_{j+1}+a_{j+1}^{\dag }b_{j}\approx \frac{1}{2}(\alpha
_{j}^{\dag }\alpha _{j+1}-\beta _{j}^{\dag }\beta _{j+1})+\mathrm{H.c.,}
\end{equation}%
neglecting the transition terms between sites with opposite potentials. Then
we have

\begin{eqnarray}
H_{0} &=&\frac{1}{2}(1-\delta )\sum_{j=1}^{N-1}\left( \alpha _{j}^{\dag
}\alpha _{j+1}-\beta _{j}^{\dag }\beta _{j+1}\right) +\mathrm{H.c.}  \notag
\\
&&+(1+\delta )\sum_{j=1}^{N}\left( \alpha _{j}^{\dag }\alpha _{j}-\beta
_{j}^{\dag }\beta _{j}\right) .
\end{eqnarray}%
The original system reduces to two independent uniform chains with opposite
chemical potentials. It can be diagonalized as%
\begin{equation}
H_{0}=\varepsilon _{k}^{0}(\alpha _{k}^{\dag }\alpha _{k}-\beta _{k}^{\dag
}\beta _{k}),
\end{equation}%
by taking the linear transformations
\begin{eqnarray}
\alpha _{k} &=&\sqrt{\frac{2}{N+1}}\sum_{j=1}^{N}(-1)^{j}\sin \frac{n\pi j}{%
N+1}\alpha _{j}, \\
\beta _{k} &=&\sqrt{\frac{2}{N+1}}\sum_{j=1}^{N}(-1)^{j}\sin \frac{n\pi j}{%
N+1}\beta _{j},
\end{eqnarray}%
where the real spectrum%
\begin{equation}
\varepsilon _{k}^{0}=(1+\delta )-(1-\delta )\cos k,
\end{equation}%
with $k=\frac{(n+1)\pi }{N+1},$ $n=0,1,2,3,...,N-1$.

Now we turn to the non-Hermitian Hamiltonian%
\begin{equation}
H=H_{0}+i\gamma \sum_{j=1}^{N}(a_{j}^{\dag }a_{j}-b_{j}^{\dag }b_{j}),
\end{equation}%
According to the result from Ref. \cite{LS17 PRA}, we still have the similar
form%
\begin{equation}
H=\sum_{k}\varepsilon _{k}(\overline{\alpha }_{k}\alpha _{k}-\overline{\beta
}_{k}\beta _{k}),
\end{equation}%
with
\begin{equation}
\varepsilon _{k}=\sqrt{\left( \varepsilon _{k}^{0}\right) ^{2}-\gamma ^{2}}.
\end{equation}%
Here the operators are defined as%
\begin{eqnarray}
\alpha _{k} &=&\sqrt{\frac{2}{N+1}}\sum_{j=1}^{N}(-1)^{j}\sin \left(
kj\right) A_{k,j}, \\
\beta _{k} &=&\sqrt{\frac{2}{N+1}}\sum_{j=1}^{N}(-1)^{j}\sin \left(
kj\right) B_{k,j},
\end{eqnarray}%
and%
\begin{eqnarray}
\overline{\alpha }_{k} &=&\sqrt{\frac{2}{N+1}}\sum_{j=1}^{N}(-1)^{j}\sin
\left( kj\right) \overline{A}_{k,j}, \\
\overline{\beta }_{k} &=&\sqrt{\frac{2}{N+1}}\sum_{j=1}^{N}(-1)^{j}\sin
\left( kj\right) \overline{B}_{k,j},
\end{eqnarray}%
where%
\begin{eqnarray}
A_{k,j} &=&\frac{a_{j}+e^{-i\varphi _{k}}b_{j}}{1+ie^{-i\varphi _{k}}}%
,B_{k,j}=\frac{a_{j}-e^{i\varphi _{k}}b_{j}}{1+ie^{-i\varphi _{k}}}, \\
\overline{A}_{k,j} &=&\frac{a_{j}^{\dag }+e^{-i\varphi _{k}}b_{j}^{\dag }}{%
1-ie^{-i\varphi _{k}}},\overline{B}_{k,j}=\frac{a_{j}^{\dag }-e^{i\varphi
_{k}}b_{j}^{\dag }}{1-ie^{-i\varphi _{k}}},
\end{eqnarray}%
and%
\begin{equation}
\tan \varphi _{k}=\frac{\gamma }{\varepsilon _{k}}.
\end{equation}%
Hamiltoinan $H$ is diagonalizable when operators $\left( \alpha _{k},\beta
_{k},\overline{\alpha }_{k},\overline{\beta }_{k}\right) $ satisfy the
canonical commutation relations
\begin{eqnarray}
\left[ \alpha _{k},\overline{\alpha }_{k^{\prime }}\right] _{\pm } &=&\left[
\beta _{k},\overline{\beta }_{k^{\prime }}\right] _{\pm }=\delta
_{kk^{\prime }},  \notag \\
\left[ \alpha _{k},\alpha _{k^{\prime }}\right] _{\pm } &=&\left[ \beta
_{k},\beta _{k^{\prime }}\right] _{\pm }=0,  \notag \\
\left[ \overline{\alpha }_{k},\overline{\alpha }_{k^{\prime }}\right] _{\pm
} &=&\left[ \overline{\beta }_{k},\overline{\beta }_{k^{\prime }}\right]
_{\pm }=0,  \notag \\
\left[ \alpha _{k},\overline{\beta }_{k^{\prime }}\right] _{\pm } &=&\left[
\overline{\alpha }_{k},\overline{\beta }_{k^{\prime }}\right] _{\pm }=0,
\notag \\
\left[ \alpha _{k},\beta _{k^{\prime }}\right] _{\pm } &=&\left[ \overline{%
\alpha }_{k},\beta _{k^{\prime }}\right] _{\pm }=0.
\end{eqnarray}%
Such relations can be directly obtained by the following relations%
\begin{eqnarray}
\left[ A_{k,j},\overline{A}_{k,l}\right] _{\pm } &=&\left[ B_{k,j},\overline{%
B}_{k,l}\right] _{\pm }=\delta _{jl},  \notag \\
\left[ A_{k,j},A_{k,l}\right] _{\pm } &=&\left[ B_{k,j},B_{k,l}\right] _{\pm
}=0,  \notag \\
\left[ \overline{A}_{k,j},\overline{A}_{k,l}\right] _{\pm } &=&\left[
\overline{B}_{k,j},\overline{B}_{k,l}\right] _{\pm }=0,  \notag \\
\left[ A_{k,j},B_{k,l}\right] _{\pm } &=&\left[ \overline{A}_{k,j},B_{k,l}%
\right] _{\pm }=0,
\end{eqnarray}%
which are always held without depending on the form of $\varphi _{k}$.

\subsection{Wavepacket trajectory}

We estimate the wavepacket trajectory in two limit cases with $\left\vert
\alpha \right\vert \ll 1$ and $\left\vert \alpha \right\vert \gg 1$. In the
first case, the initial state is approximately expressed as
\begin{equation}
\left\vert \psi (0)\right\rangle =\frac{1}{\sqrt{1+\left\vert \alpha
\right\vert ^{2}}}(\left\vert 0\right\rangle +\alpha \left\vert
1\right\rangle ).
\end{equation}%
so
\begin{eqnarray}
\left\vert \psi (t)\right\rangle &=&U(t)\left\vert \psi (0)\right\rangle
\notag \\
&=&\frac{1}{\sqrt{1+\left\vert \alpha \right\vert ^{2}}}e^{-i\omega t}\left(
\left\vert 0\right\rangle +\alpha e^{-i\omega t}\left\vert 1\right\rangle
\right) ,
\end{eqnarray}%
where
\begin{equation}
\left\vert n\right\rangle =\sqrt{\frac{2}{1+N}}\sum_{j=1}^{N}\sin \frac{%
(n+1)\pi j}{N+1}\left\vert j\right\rangle ,
\end{equation}%
then we have
\begin{eqnarray}
&&r_{c}\left( t\right) =\sum_{l=1}^{N}l\left\vert \left\langle l\right\vert
\psi (t)\rangle \right\vert ^{2}  \notag \\
&=&-\frac{8\left\vert \alpha \right\vert \cos (\frac{\pi }{1+N})\cot ^{2}(%
\frac{\pi }{2+2N})}{(1+\left\vert \alpha \right\vert ^{2})(1+N)[1+2\cos (%
\frac{\pi }{1+N})]^{2}}\cos (\omega t-\arg \alpha )  \notag \\
&&+\frac{1}{2}\left( 1+N\right) ,
\end{eqnarray}%
in the small $\left\vert \alpha \right\vert $ and large $N$ limit, by using
\begin{equation}
\lim_{N\rightarrow \infty }\frac{2\cos (\frac{\pi }{1+N})\cot ^{2}(\frac{\pi
}{2+2N})}{(1+N)^{2}[1+2\cos (\frac{\pi }{1+N})]^{2}}=\frac{8}{9\pi ^{2}},
\end{equation}%
we get the wavepacket trajectory

\begin{equation}
r_{c}\left( t\right) \approx -\frac{32N}{9\pi ^{2}}\left\vert \alpha
\right\vert \cos (\omega t-\arg \alpha )+\frac{N}{2}.
\end{equation}%
Secondly, in the large $\alpha =\left\vert \alpha \right\vert $\ limit, we
have
\begin{equation}
\left\vert \alpha (t)\right\rangle \approx \left( \frac{1}{\alpha \sqrt{2\pi
}}\right) ^{1/2}\sum_{n=0}^{N-1}e^{-\left( n-\alpha ^{2}\right) ^{2}/4\alpha
^{2}}e^{-i(n+1)\omega t}\left\vert n\right\rangle .
\end{equation}%
The wavepacket trajectory is%
\begin{equation}
r_{c}(t)=\sum_{j=1}^{N}jP\left( t,j\right) ,
\end{equation}%
with
\begin{equation}
P\left( t,j\right) \approx \left\vert \sum_{n=0}^{N-1}\frac{2e^{-\left(
n-\alpha ^{2}\right) ^{2}/4\alpha ^{2}}e^{-in\omega t}}{\sqrt{\alpha 2N\sqrt{%
2\pi }}}\sin (\frac{n\pi j}{N})\right\vert ^{2}.
\end{equation}%
The oscillating terms have no contribution to the summation, leading to
\begin{equation}
P\left( t,j\right) =\left\{
\begin{array}{c}
1,\text{ }j=\pm \frac{N\omega t}{\pi } \\
0,\text{otherwise}%
\end{array}%
\right. .
\end{equation}%
Then we have
\begin{equation}
r_{c}(t)=\frac{N\left( \omega t-2n\pi \right) }{\pi }\left\{
\begin{array}{c}
1,\text{ }t\in \lbrack 0,\frac{\pi }{\omega })+\frac{2n\pi }{\omega } \\
-1,\text{ }t\in \lbrack -\frac{\pi }{\omega },0)+\frac{2n\pi }{\omega }%
\end{array}%
,n\in Z\right. ,
\end{equation}%
which represents a tranglar wave with period $\frac{2\pi }{\omega }$.

\acknowledgments We acknowledge the support of the CNSF (Grant No. 11374163).


\begin{thebibliography}{99}
\bibitem{Griffiths} D. J. Griffiths, \textit{Introduction to Quantum
Mechanics} (2nd ed.), Prentice Hall, ISBN 0-13-805326-X (2004).

\bibitem{Liboff} R. L. Liboff, \textit{Introductory Quantum Mechanics},
Addison--Wesley, ISBN 0-8053-8714-5 (2002).

\bibitem{Zeng} J.-Y. Zeng, \textit{Quantum Mechanics} Vol. I (in Chinese),
3rd Edition, Science Press, Beijing, (2006).

\bibitem{Glauber} J. Glauber, Phys. Rev. \textbf{131,} 2766 (1963).

\bibitem{Bender 98} C. M. Bender and S. Boettcher, Phys. Rev. Lett. \textbf{%
80}, 5243 (1998).

\bibitem{Bender 99} C. M. Bender, S. Boettcher, and P. N. Meisinger, J.
Math. Phys. \textbf{40,} 2201 (1999).

\bibitem{Dorey 01} P. Dorey, C. Dunning, and R. Tateo, J. Phys. A: Math.
Gen. \textbf{34,} L391 (2001).

\bibitem{Dorey 02} P. Dorey, C. Dunning, and R. Tateo, J. Phys. A: Math. Gen
\textbf{34,} 5679 (2001).

\bibitem{Bender 02} C. M. Bender, D. C. Brody, and H. F. Jones, Phys. Rev.
Lett. \textbf{89,} 270401 (2002).

\bibitem{A.M43} A. Mostafazadeh, J. Math. Phys. \textbf{43,} 3944 (2002).

\bibitem{A.M36} A. Mostafazadeh, J. Phys. A: Math. Gen. \textbf{36,} 7081
(2003).

\bibitem{Jones} H. F. Jones, J. Phys. A: Math. Gen. \textbf{38,} 1741 (2005).

\bibitem{M.Z40} M. Znojil, J. Phys. A \textbf{40,} 13131 (2007).

\bibitem{M.Z41} M. Znojil, J. Phys. A \textbf{41,} 292002 (2008).

\bibitem{M.Z82} M. Znojil, Phys. Rev. A \textbf{82,} 052113 (2010).

\bibitem{AGuo} A. Guo. Phys. Rev. Lett. \textbf{103,} 093902 (2009).

\bibitem{CERuter} C. E. R\"{u}ter, et al. Nat. Phys. \textbf{6,} 192-195
(2010).

\bibitem{Wan} W. Wan, Y. Chong, L. Ge, H. Noh, A. D. Stone, and H. Cao,
Science \textbf{331,} 889-892 (2011).

\bibitem{Sun} Y. Sun, W. Tan, H.-Q. Li, J. Li, and H. Chen, Phys. Rev. Lett.
\textbf{112,} 143903 (2014).

\bibitem{LFeng} L. Feng, et al. Nature Mater. \textbf{12,} 108-113 (2013).

\bibitem{BPeng} B. Peng, et al. Nat. Phys. \textbf{10,} 394-398 (2014).

\bibitem{LChang} L. Chang, et al. Nature Photon. \textbf{8,} 524-529 (2014).

\bibitem{LFengScience} L. Feng, Z. J. Wong, R.-M. Ma, Y. Wang, and X. Zhang,
Science \textbf{346,} 972-975 (2014).

\bibitem{HodaeiScience} H. Hodaei, et al. Science \textbf{346,} 975-978
(2014).

\bibitem{NC2015} M. Wimmer, et al. Nat. Commun \textbf{6,} 7782 (2015).

\bibitem{A.M} A. Mostafazadeh and A. Batal, J. Phys. A: Math. Gen. \textbf{%
	37,} 11645 (2004).

\bibitem{A.M38} A. Mostafazadeh, J. Phys. A: Math. Gen. \textbf{38,} 6557
(2005).

\bibitem{A.M391} A. Mostafazadeh, J. Phys. A: Math. Gen. \textbf{39,} 10171
(2006).

\bibitem{A.M392} A. Mostafazadeh, J. Phys. A: Math. Gen. \textbf{39}, 13495
(2006).

\bibitem{JLPT} L. Jin and Z. Song, Phys. Rev. A \textbf{80,} 052107 (2009).

\bibitem{L.J81} L. Jin and Z. Song, Phys. Rev. A \textbf{81,} 032109 (2010).

\bibitem{L.J83} L. Jin and Z. Song\textbf{, }Phys. Rev. A \textbf{83,}
062118 (2011).

\bibitem{L.J44} L. Jin and Z. Song, J. Phys. A: Math. Theor. \textbf{44,}
375304 (2011).

\bibitem{LJin84} L. Jin and Z. Song, Phys. Rev. A \textbf{84}, 042116 (2011).

\bibitem{ST} T. Shi, Y. Li, Z. Song, and C.P. Sun, Phys. Rev. A \textbf{71},
032309 (2005); T. Boness, S. Bose, and T.S. Monteiro, Phys. Rev. Lett. 96
187201 (2006).

\bibitem{SSH} W. P. Su, J. R. Schrieffer, and A. J. Heeger, Phys. Rev. Lett.
\textbf{42,} 1698 (1979).

\bibitem{Zak} J. Zak, Phys. Rev. Lett. \textbf{62,} 2747 (1989).

\bibitem{Asboth} J. K. Asb\'{o}th, L. Oroszl\'{a}ny, and A. P\'{a}lyi,
\textit{A Short Course on Topological Insulators: Band Structure and Edge
States in One and Two Dimensions, }Lecture Notes in Physics\textit{\ }
(Springer International Publishing, Switzerland, 2016).

\bibitem{HWH} W. H. Hu, L. Jin, Y. Li, and Z. Song, Phys. Rev. A \textbf{86}
, 042110, (2012).

\bibitem{LS17 PRA} S. Lin, and Z. Song, Phys. Rev. A \textbf{96}, 052121
(2017).
\end{thebibliography}
\end{document}